\documentclass[preprint,11pt]{aastex}
\usepackage[utf8]{inputenc}
\usepackage{pdflscape}
\usepackage{geometry}
\usepackage{amsmath, amsthm, amssymb}
\usepackage{natbib}

\title{The APOKASC Catalog: An Asteroseismic and Spectroscopic Joint Survey of Targets in the Kepler Fields}

\author{Marc H. Pinsonneault\altaffilmark{1,2}, Yvonne Elsworth\altaffilmark{3,4}, Courtney Epstein\altaffilmark{1}, Saskia Hekker\altaffilmark{5}, Sz. M\'esz\'aros\altaffilmark{6}, 
 William J. Chaplin\altaffilmark{3,4}, Jennifer A. Johnson\altaffilmark{1,2}, Rafael A. Garc\'ia\altaffilmark{7}, Jon Holtzman\altaffilmark{8}, Savita Mathur\altaffilmark{9}, 
 Ana Garc\'{\i}a P\'erez\altaffilmark{10}, Victor Silva Aguirre\altaffilmark{4}, L\'eo Girardi\altaffilmark{11,12}, Sarbani Basu\altaffilmark{13}, Matthew Shetrone\altaffilmark{14}, Dennis Stello\altaffilmark{4,15}, 
Carlos Allende Prieto\altaffilmark{16,17}, Deokkeun An\altaffilmark{18}, Paul Beck\altaffilmark{7}, Timothy C. Beers\altaffilmark{19,20}, Dmitry Bizyaev\altaffilmark{21}, Steven Bloemen\altaffilmark{22}, 
Jo Bovy\altaffilmark{23}, Katia Cunha\altaffilmark{24,25}, Joris De Ridder\altaffilmark{26}, Peter M. Frinchaboy\altaffilmark{27},
D.A. Garcia-Hern\'andez\altaffilmark{16,17}, Ronald Gilliland\altaffilmark{28}, Paul Harding\altaffilmark{29}, Fred R. Hearty\altaffilmark{28}, Daniel Huber\altaffilmark{30,31}, Inese Ivans\altaffilmark{32}, 
Thomas Kallinger\altaffilmark{33},
Steven R. Majewski\altaffilmark{10}, Travis S. Metcalfe\altaffilmark{9}, Andrea Miglio\altaffilmark{3,4}, Benoit Mosser\altaffilmark{34}, Demitri Muna\altaffilmark{1},
David L. Nidever\altaffilmark{35}, Donald P. Schneider\altaffilmark{28,36},
Aldo Serenelli\altaffilmark{37}, Verne V. Smith\altaffilmark{38}, Jamie Tayar\altaffilmark{1}, Olga Zamora\altaffilmark{16,17},Gail Zasowski\altaffilmark{39}}

\altaffiltext{1}{Dept. of Astronomy, The Ohio State University, Columbus, OH 43210, USA}
\altaffiltext{2}{Center for Cosmology and Astroparticle Physics, The Ohio State University, Columbus OH, 43210 USA}
\altaffiltext{3}{University of Birmingham, School of Physics and Astronomy, Edgbaston, Birmingham B15 2TT, UK}
\altaffiltext{4}{Stellar Astrophysics Centre, Department of Physics and Astronomy, Aarhus University, Ny Munkegade 120, DK-8000 Aarhus C, Denmark}
\altaffiltext{5}{Max-Planck-Institut fur Sonnensystemforschung, Justus-von-Liebig-Weg 3, 37077 Gottingen, Germany}
\altaffiltext{6}{Astronomy Department, Indiana University, Bloomington,IN 47405, USA}
\altaffiltext{7}{Laboratoire AIM, CEA/DSM-CNRS - Universit\'e Denis Diderot-IRFU/SAp, 91191, Gif-sur-Yvette Cedex, France}
\altaffiltext{8}{Department of Astronomy, MSC 4500, New Mexico State University, P.O. Box 30001, Las Cruces, NM 88003, USA} 
\altaffiltext{9}{Space Science Institute, 4750 Walnut street Suite 205, Boulder, CO 80301, USA}
\altaffiltext{10}{Department of Astronomy, University of Virginia, P.O.Box 400325, Charlottesville, VA 22904-4325, USA}
\altaffiltext{11}{Osservatorio Astronomico di Padova -- INAF, Vicolo dell'Osservatorio 5, I-35122 Padova, Italy}
\altaffiltext{12}{Laborat\'orio Interinstitucional de e-Astronomia -- LIneA, Rua Gal. Jos\'e Cristino 77, Rio de Janeiro, RJ -- 20921-400, Brazil}
\altaffiltext{13}{Department of Astronomy, Yale University, PO Box 208101, New Haven, CT 06520-8101, USA}
\altaffiltext{14}{University of Texas at Austin, McDonald Observatory 32 Fowlkes Rd. McDonald Observatory, TX 79734-3005, USA}
\altaffiltext{15}{Sydney Institute for Astronomy (SIfA), School of Physics, University of Sydney, NSW 2006, Australia}
\altaffiltext{16}{Instituto de Astrofsica de Canarias (IAC), C/Va Lactea, s/n, E-38200, La Laguna, Tenerife, Spain}
\altaffiltext{17}{Departamento de Astrofsica, Universidad de La Laguna, E-38206, La Laguna, Tenerife, Spain}
\altaffiltext{18}{Department of Science Education, Ewha Womans University, Seoul, Korea}
\altaffiltext{19}{Department of Physics, University of Notre Dame, 225 Nieuwland Science Hall, Notre Dame, IN 46656, USA}
\altaffiltext{20}{JINA: Joint Institute for Nuclear Astrophysics, University of Notre Dame, Notre Dame, IN 46556, USA}
\altaffiltext{21}{Apache Point Observatory and New Mexico State University, P.O. Box 59, Sunspot, NM, 88349-0059, USA}
\altaffiltext{22}{Department of Astrophysics, IMAPP, Radboud University Nijmegen, PO Box 9010, NL-6500 GL Nijmegen, The Netherlands}
\altaffiltext{23}{Institute for Advanced Study, Einstein Drive, Princeton, NJ 08540, USA}
\altaffiltext{24}{Observat\'orio Nacional, Sao Crist\'ov\~ao, Rio de Janeiro, Brazil}
\altaffiltext{25}{Steward Observatory, University of Arizona, Tucson, AZ 85719, USA}
\altaffiltext{26}{Instituut voor Sterrenkunde, KU Leuven, 3001, Leuven, Belgium}
\altaffiltext{27}{Department of Physics \& Astronomy, Texas Christian University, Fort Worth, TX 76129, USA}
\altaffiltext{28}{Department of Astronomy and Astrophysics, The Pennsylvania State University, University Park, PA 16802, USA}
\altaffiltext{29}{Department of Astronomy, Case Western Reserve University, Cleveland, OH 44106-7215, USA}
\altaffiltext{30}{NASA Ames Research Center, Moffett Field, CA 94035, USA}
\altaffiltext{31}{SETI Institute, 189 Bernardo Avenue, Mountain View, CA 94043, USA}
\altaffiltext{32}{Department of Physics and Astronomy, The University of Utah, Salt Lake City, UT 84112, USA}
\altaffiltext{33}{Institute for Astronomy, University of Vienna, T\"urkenschanzstrasse 17, 1180 Vienna, Austria}
\altaffiltext{34}{LESIA, UMR 8109, Universit\'e Pierre et Marie Curie, Universit\'e Denis Diderot, Observatoire de Paris, 92195 Meudon Cedex, France}
\altaffiltext{35}{Department of Astronomy, University of Michigan, Ann Arbor, MI 48104 USA}
\altaffiltext{36}{Institute for Gravitation and the Cosmos, The Pennsylvania State University, University Park, PA 16802, USA}
\altaffiltext{37}{Institute of Space Sciences (IEEC-CSIC), Campus UAB, 08193, Bellaterra, Spain}
\altaffiltext{38}{National Optical Astronomy Observatories, Tucson, AZ 85719 USA}
\altaffiltext{39}{Department of Physics \& Astronomy, Johns Hopkins University, Baltimore, MD, 21218, USA}

\def\dnu{$\Delta\nu$}
\def\numax{$\nu_{\rm max}$}

\def\logg{$\log g$}

\input epsf

\begin{document}

\begin{abstract}

We present the first APOKASC catalog of spectroscopic and asteroseismic properties of 1916 red giants observed in the \emph{Kepler} fields.
The spectroscopic parameters provided from the Apache Point Observatory Galactic Evolution Experiment project are complemented 
with asteroseismic surface gravities, masses, radii, and mean densities determined by members of the \emph{Kepler} Asteroseismology Science Consortium.  
We assess both random and systematic sources of error 
and include a discussion of sample selection for giants in the \emph{Kepler} fields.
 Total uncertainties in the main catalog properties are of order 80 K in $T_{\rm eff}$, 0.06 dex in [M/H], 0.014 dex in $\log g$, and $12 \%$
and $5 \%$  in mass and radius, respectively; these reflect a combination of systematic and random errors.  
Asteroseismic surface gravities are substantially more precise and accurate than spectroscopic ones, 
and we find good agreement between their mean values and the calibrated spectroscopic surface gravities.  There are, however, 
systematic underlying trends with $T_{\rm eff}$ and $\log g$.  Our effective temperature scale is between 0-200 K cooler than that expected from the Infrared Flux Method, 
depending on the adopted extinction map,
which provides evidence for a lower value on average than that inferred for the Kepler Input Catalog (KIC).  We find a reasonable 
correspondence between the photometric KIC and spectroscopic APOKASC metallicity scales, with increased dispersion in KIC metallicities
as the absolute metal abundance decreases, and offsets in $T_{\rm eff}$ and $\log g$ consistent with those derived in the literature.  
We present mean fitting relations between APOKASC and KIC observables and discuss future prospects, strengths, and limitations of the catalog data.

\end{abstract}

\section{Introduction}

We are entering the era of precision stellar astrophysics.  Large surveys are yielding data 
with unprecedented quality and quantity, and even more ambitious programs are on the near horizon.  
This advance is not merely a matter of much larger samples of measurements than were 
possible before; there are also fundamentally new observables arising from the advent of 
asteroseismology as a practical stellar population tool.  These new observables are particularly powerful
diagnostics when complemented with data from more traditional approaches.  In this paper we present the first 
release of the joint APOKASC asteroseismic and spectroscopic survey for targets with both high-resolution 
Apache Point Observatory Galactic Evolution Experiment (APOGEE) spectra
analyzed by members of the third Sloan Digital Sky Survey (SDSS-III) and asteroseismic data obtained by the \emph{Kepler} mission and
analyzed by members of the \emph{Kepler} Asteroseismology Science Consortium (KASC).  When completed we
anticipate  of order 8,000 red giants and 600 dwarfs and subgiants with asteroseismic data and high-resolution spectra. 
In this initial paper we catalog the properties of 1916 red giants observed as part of the Sloan Digital Sky Survey Data Release 10 \citep{Ahn14}.  
A catalog for the less-evolved stars will be presented in a separate publication (Serenelli et al. 2014, in prep.)

Large spectroscopic surveys in the Milky Way galaxy are now a reality, and a variety of sampling strategies and resolutions have been employed.  
Low- to medium-resolution surveys
such as SEGUE \citep{Yanny09}, RAVE \citep{Kordopatis13}, and LAMOST \citep{Zhao06} provide stellar properties for large samples of stars.  
High-resolution programs are complementary, with more detailed abundance mixtures and more precise measurements for still-substantial samples.
GALAH \citep{Freeman12} and Gaia-ESO \citep{Gilmore12} are optical surveys; APOGEE \citep{Majewski10,Hayden14} is instead focusing
on the infrared. Infrared spectroscopy is attractive for Milky Way studies because it is less sensitive to extinction; it also has different systematic error sources than
traditional optical spectroscopy (Garc\'{\i}a P\'erez 2014, in prep.) These surveys permit detailed stellar population reconstructions using chemical tagging, kinematic data, effective temperature, 
and surface gravities; see for example, \citet{Bovy12} for SEGUE, \citet{Bergemann14} for Gaia-ESO, or \citet{Binney14} for RAVE.  Spectroscopic properties can be complemented by photometric parameter estimation, 
and the Gaia mission \citep{Perryman01} should add
critical measurements of distances and proper motions.  

Standing alone, however, there are intrinsic limitations in the information from spectroscopic studies of stellar populations. The fundamental stellar properties of
mass and age are only indirectly inferred from spectra.  In the case of red giants, their HR diagram position yields relatively weak constraints on either mass or age.  Chemical tagging \emph{-} for 
example, using high [$\alpha$/Fe] as a marker of old populations \citep{Wallerstein62,Tinsley79} \emph{-} is a valuable method, but the absolute time scale is uncertain \citep{Matteucci09} and the chemical
 evolution rates in different systems need not have been the same.   Large surveys also require automated pipeline estimation of stellar parameters, and it
can be difficult to calibrate these pipelines. These issue reflect the underlying problem that the
dependence of absorption line strength on stellar atmosphere properties is not completely understood.  This is in part due to incomplete and inaccurate
atomic data, but there are also important physical effects that are challenging to model.  Traditional atmosphere analysis uses one-dimensional plane-parallel 
atmospheres, treats line broading with an ad hoc microturbulence, and assumes local thermodynamic equilibrium (LTE).  Spherical effects, more realistic turbulence modeling from three dimensional \
hydrodynamic simulations, and departures from LTE can strongly impact the interpretation of the spectrum (see for example \citealt{Asplund05}).  The difficulty of
modeling the outermost layers, where real atmospheres transition
to a chromosphere, can also play a role in complicating spectroscopic inferences. Therefore, stellar parameter determinations
for large samples require calibration against standards whose properties
are known by other means, or whose membership in a cluster demands that
they have similar composition.

Independently, there have been exciting advances in asteroseismology driven largely by data from space missions.  In an important breakthrough, the 
CoRoT \citep{DeRidder09} and \emph{Kepler} \citep{Bedding10} missions have discovered that virtually all red giants are non-radial oscillators. 
Red giant asteroseismology is revolutionizing our understanding of stellar structure and evolution. Global properties of the oscillations, such as the large frequency 
spacing $\Delta\nu$ and frequency of maximum oscillation power $\nu_{\rm max}$, are naturally related to the stellar mean density and surface gravity, respectively \citep{Kjeldsen95};
see also \citet{Stello09b} and \citet{Huber11}.   
Basic pulsation data, combined with measured effective temperatures, permit estimation of stellar masses and radii in a domain 
where these crucial stellar properties have been notoriously
 uncertain.  Access to large numbers of stellar mass measurements has profound implications for stellar population studies \citep{Miglio09,Freeman11}.  
Seismology also yields completely new observables 
for red giants because of a fortunate coincidence: their physical structure permits coupling between waves that propagate primarily in the core and those which 
propagate primarily in the envelope. The net result is oscillations of mixed character which carry information about the structure of both the core and envelope.
We can therefore use the detailed pattern of observed oscillation frequencies to distinguish between first-ascent giants (with H-shell burning only) and red clump stars (with He-core and H-shell burning) \citep{Bedding11}.  
Asymptotic red giant stars, with two shell sources, would appear with a pattern similar, but not identical, to first-ascent giants; 
however, most work to date has focused on lower luminosities where such stars are not expected to be found.
Rapid rotation in the cores of red giants has been discovered 
(\citealt{Beck12,Deheuvels12,Mosser12a}).  Time-series data from \emph{Kepler} can also measure the surface rotation rates of stars through 
starspot modulation of their light curves \citep{Basri11,Garcia14b,McQuillan14}.  Mapping the angular momentum evolution of giants as a function of mass is another 
new frontier with rich astrophysical rewards.

Asteroseismology alone, however, has important limitations on the information that it can provide.  Effective temperatures are 
required to infer mass and radius separately, and stellar ages depend on both mass and composition.
For example, it was possible to use the Kepler Input Catalog (KIC) data of \citet{Brown11}
to define a sequence of solar-mass asteroseismic targets from the main
sequence to the giant branch, but abundances are essential for finding true solar analogs \citep{SilvaAguirre11}.  In the CoRoT fields \citet{Miglio13} were able to 
infer mean stellar mass differences between populations along different sightlines, but interpreting these measurements in terms of age would require spectroscopic data on
metallicity and the mixture of heavy elements.  The scarcity of abundance data for stars in the \emph{Kepler} field is therefore a major limitation; 
the sheer volume of data (roughly 21,000 red giants) has made 
traditional spectroscopic studies infeasible.

Fortunately, there is a new spectroscopic survey ideally suited for large samples of red giants. The multi-fiber,
 high-resolution H-band spectrograph from APOGEE on the SDSS 2.5 m telescope \citep{Gunn06}, is ideally suited to observing \emph{Kepler} targets because it is well 
matched to the target density of the fields observed by the mission
with 230 science fibers available over a 7 square degree field. The $R=22,500$ spectra were designed to produce temperatures, [Fe/H] and [X/Fe] 
with accuracies of $4\%$, 0.1 dex, and 0.1 dex respectively \citep{Eisenstein11}.  Actual performance, as reported by \citet{Meszaros13}, is close to these goals.
There is no other spectroscopic sample of this size and quality available for \emph{Kepler} red giants.
\footnote{The APOGEE data used in this paper adopted for an abundance scale the metallicity index [A/H], which corresponds to the metallicity [Fe/H] (see Section 3.1.)
Future APOGEE data releases will provide abundance measurements of 15 elements, including O, Mg, and Fe.}

Here we report the first APOKASC data release, which includes red giants whose spectra were released in the SDSS-III Date Release 10 (DR10), as
described in \citet{Ahn14}.
Our paper is organized as follows: We describe our sample selection in Section 2.  Our spectroscopic data calibration is described 
and compared with photometric temperatures and asteroseismic surface gravities in Section 3.  The asteroseismic analysis is discussed in Section 4.  
We present the catalog and compare it to the KIC in Section 5.  A look ahead to the full catalog and a discussion is presented in Section 6.

\section{Sample Selection}

Our goal is to obtain data for a combined asteroseismic and spectroscopic sample of a large number of astrophysically interesting targets in the \emph{Kepler} fields. Understanding
the selection effects in our sample is important for interpreting our results.  Selection effects in our sample enter
at several distinct levels: the match between the oscillation frequencies and the time sampling, the selection of which
stars to study for oscillations, and the selection of targets for spectroscopic observation.

The oscillation frequencies span a broad range, from five minutes 
for the Sun to tens of days for luminous giants.  A single observing strategy will therefore not work 
across the entire domain.  Fortunately, \emph{Kepler} has two observing modes: one minute (short-cadence) 
and thirty minute (long-cadence).  Long-cadence targets are typically observed in 90 day cycles, hereafter referred to as quarters. 
The short-cadence mode is ideal for asteroseismic studies of 
dwarfs and subgiants; however, there are a limited number of such targets that \emph{Kepler} was able to 
observe.  \citet{Chaplin11} detected oscillations in $\sim$ 600 of the $\sim$ 2,000 targets observed in short-cadence mode for at least 30 days.  The long-cadence 
mode is ideal for measuring oscillations in red giants, and a large number 
 were observed for at least one quarter by the satellite.  More than 70\% of the 
long-cadence red giants were detected as solar-like oscillators even with the first three months of
data \citep{Hekker11a}, with the non-detections primarily in luminous stars requiring a longer time
sequence.  Furthermore, the most precise measurements derived from IR spectra are those for cool and evolved 
stars, so there is a natural pairing between APOGEE spectra and asteroseismology of red giants.  The bulk 
of our sample is therefore composed of red giants, with smaller, separate designated dwarf and subgiant cohorts,
and the duration of the mission and time sampling should not introduce signficant biases in our sample.  

However, there are more bright red giants in the \emph{Kepler} fields than were observed in long-cadence mode,
and there are more long-cadence targets than the number for which it was practical to obtain spectra.  There are 
therefore two distinct selection criteria important for this sample: the criteria for being observed by \emph{Kepler} and the criteria for being observed 
spectroscopically by APOGEE.  Furthermore, the initial DR10 sample is a subset of the overall 
APOKASC sample, and the properties of these fields need not be representative for the sample as a whole.  
We therefore begin with a summary of the \emph{Kepler} short and long-cadence target selection procedures.
We then describe how we used preliminary asteroseismic data to identify populations for spectroscopic observation.  We then discuss special populations which we targeted for observing,
 describe our main sample grid, and how we filled the remaining fibers for the campaign.   We end with a brief discussion of 
the properties of the first dataset being released in this paper.

\subsection {Kepler target selection}

We included 400 of the asteroseismically-detected subgiants and dwarfs reported in \citet{Chaplin11} from short-cadence observations.  This sample is
smaller than that of \citet{Chaplin11} because hotter dwarfs did not fit our global criteria
for APOGEE observations.  The selection process of targets from the long-cadence sample is more complex than that for the 
short-cadence sample.  
The Kepler mission was designed to search for transits of host stars by extrasolar planets, with a focus on solar analogs.  However, giants are much more numerous than
dwarfs in a sample with the \emph{Kepler} magnitude limit, which was designed for stars brighter than Kepler magnitude \emph{$K_p$} = 16.  
The Kepler Input Catalog (KIC; \citealt{Brown11}) was therefore constructed to separate
dwarfs and giants and to define the planet candidate target list.  There were multiple criteria used to select giants for long-cadence observations.  For
our purposes there are two important samples: the KASC giants, defined below, and the full sample (hereafter referred to as the public giants).
  
A pre-launch list of 1,006 targets was chosen using only information available from the KIC, and with the express purpose of providing 
a uniformly spaced set of stars over the focal plane serving as low proper motion, small parallax, astrometric controls.  
As such they were selected on the basis of a metric that was a combination of:  (a) large distance,
(b) bright, but not expected to saturate the detector to allow precise centroiding, (c) uncrowded -- also to support precise centroiding, 
and (d) spread over the focal plane to give 11 to 12 red giant controls for each of \emph{Kepler's} 84 channels.  
There was also a sample
of $\sim$ 800 giants for asteroseismic monitoring that was assembled from pre-launch proposals submitted by KASC working
groups.  \footnote{A full list of proposals can be accessed on the KASOC database.} The initial red giant asteroseimology results 
were based on the combination of these two datasets; hereafter we refer to these targets as the KASC giants.  \citet{Huber10} described how the 
basic seismic parameters were measured; \citet{Kallinger10} provided the stellar parameters, such as mass and radius; and \citet{Hekker11b} 
compared different analysis techniques.  However, the derived stellar parameters did not include high-resolution spectroscopic metallicities and 
effective temperature estimates independent of the KIC, which APOGEE can now provide.

Characterizing the full public red giant sample is surprisingly challenging. The target list varied from observing quarter to quarter, as the giants
candidates were deprioritized relative to the dwarfs for planet searches.  Using data from the first sixteen quarters in the Q1-Q16 star properties catalog, \citet{Huber14} combined the KIC and published 
literature information to obtain a total of 
21,427 stars with $\log g$ $< 3.5$ and $T_{\rm eff} < 5500K $ that were observed for at least one quarter during the \emph{Kepler} mission. 
This sample does not include stars observed during commissioning (Q0) only.

The majority of the \citet{Huber14} sample, of order 15,000 red giants, were selected as planet search candidates using the procedure described in \citet{Batalha10}.  
A total of 5282 red giants brighter than \emph{$K_p$} = 14 were included in the highest priority
planet search cohort of 150,000 stars.  The remaining red giants 
were selected from a much larger secondary target list of 57,010 giants, including a large number (11,057) brighter than \emph{$K_p$} = 14.
In practice this
criterion favored brighter targets that were classifed as smaller red giants; this is roughly equivalent to a sample with magnitude and
surface gravity cuts.
The mission also supplemented the target list with a cohort of \~12,000 brighter targets, including stars without KIC classification;
$\sim$ 3,300 of these unclassfied stars proved to be cool or luminous giants \citep{Huber14}.  Approximately 1,000 giant targets were also added in the GO
program, including $\sim$ 300 that were observed in Quarters 14-16 as part of a dedicated APOKASC GO proposal $\#40033$ as described below.  There are also
an unknown number of stars classified as dwarfs in the KIC that could be giants, and vice versa.  An asteroseismic luminosity classification of all cool
dwarfs is planned and should yield a complete census of giant stars in 2014.

\subsection {Target Selection for the Overall APOKASC Sample}

Observing the entire \emph{Kepler} red giant sample with APOGEE was not feasible, so APOKASC had to develop an independent spectroscopic target selection process. There are high-priority
categories of targets where we attempted to be as complete as possible \emph{-} for example, rare but astrophysically interesting metal-poor stars or open
cluster members.  We also wanted uniform spectroscopic data for the stars with the highest quality \emph{Kepler} light curves, as these could serve as precise
calibrators for stellar population and asteroseismic studies.  Another survey goal was to extend the range of surface gravity and metallicity relative to prior 
spectroscopic studies.
Finally, ages can be inferred from masses for first-ascent red giants, so it was important to have estimates of evolutionary state to preferentially target
such stars over the more common core-He burning stars where mass loss complicates the mapping from mass to age.  

Our procedures for creating the final target list is described below.  For a further discussion of APOGEE and APOKASC targeting, see
\citet{Zasowski13}.  For all targets we adopted limits on the magnitude ($7 < H < 11$) and 
effective temperature ($T_{\rm eff} < 6500$ K) that were necessary for APOGEE, and we required that the targets fit within the APOGEE field of view.  The magnitude limits guarded against overexposure and
ensured a high signal-to-noise ratio in 1 hour observations; the temperature cut was designed to avoid hot stars with uninformative IR spectra.  
We performed a uniform analysis of the \emph{Kepler} light curves to check for evolutionary diagnostics,
rotation, or unusual asteroseismic properties.  We added in external information for interesting populations.  We then defined
a reference set of red giants and red clump stars sampling a wide range of surface gravities, prioritizing stars with more complete time coverage.
These datasets left us with free fibers that we could fill from the remainder of the available giant sample.  

\subsubsection {Asteroseismic Classification}

We employed asteroseismic diagnostics for the full \emph{Kepler} red giant sample observed in long-cadence mode to infer their evolutionary state 
and to identify stars with unusual pulsational properties.
We discuss the methods used for extracting the basic asteroseismic observables for the catalog in Section 4.  Our automated methodology for evolutionary state classification is described 
in \citet{Stello13} and summarized here.  The key tasks are identifying the frequencies of the dipole (angular degree $l = 1$)
oscillation modes and inferring whether the pattern is characteristic of a core He-burning star or one with a degenerate and inert He core.  To obtain the frequencies
we detrended the time series from the public Q1 to Q8 data by removing discontinuities and applying a high-pass filter.  The SYD pipeline \citep{Huber09} was used to derive the large frequency separation, and we 
implemented a simple peak bagging approach using different degrees of smoothing to identify all significant peaks in the power spectra. We associate a degree $l$ to each extracted frequency based on a method 
similar to that provided by \citet{Mosser11a}. For the purpose of measuring period spacings of the dipole modes, which is important for this classification step, we remove the radial and quadrupole modes. 

We only kept 
stars for which we detected at least five $l = 1$ modes in order to obtain a more robust result in the following step. There are more than 8,000 stars that pass this criterion. For each star we measure 
the pairwise period spacing, $\Delta P$, between successive peaks, and take the median of $\Delta P$ as the representative period spacing (the median proved to be the most robust quantity compared 
to a simple mean or the moment). This method produced our best diagnostic of evolutionary state.  We complemented this approach with a double-check using the methodology of \citet{Mosser11b}.  
A subset of our targets (3128) could be unambiguously assigned to either the red clump or the first-ascent red giant branch.  Limited frequency resolution and other backgrounds, such as rotation, made automated
classification ambiguous for the remainder of the sample.  This evolutionary state information was used only for target selection
purposes, as our procedure was not designed to provide complete information for the entire sample.

\subsubsection {KASC Giants, Open Cluster Members, Asteroseismic Dwarfs and Subgiants}

We began our program by defining the highest-priority targets.  The KASC giants described above are the best studied stars with the highest quality datasets, so we ensured that
all of them would be included as targets.  All subgiants and dwarfs from \citet{Chaplin11} with KIC $T_{\rm eff} < 6500 K$ and asteroseismic detections were targeted.  
Known members of the open clusters NGC 6811, 6819, and 6791 bright enough for APOGEE and with asteroseismic detections were also all included \citep{Stello11a}. We also
developed criteria to preferentially select metal-poor giants, rapid rotators, and luminous giants for spectroscopic observation.

\subsubsection {Metal-Poor Giants}

Observing oscillations in metal-poor giants is critical for both stellar physics (e.g., scaling relations as a function of metallicity; see \citealt{Epstein14}) and stellar populations questions, such as 
the age of the halo.  A simulation of the stellar populations in the \emph{Kepler} fields with the TRILEGAL code \citep{Girardi05} 
indicated that only 0.6\% of giants are expected to have $[Fe/H] < -2.0$, for a predicted total of $\sim 100$ in the entire public giant database. Efficient targeting of these stars is therefore essential,  
and we employed several methods to identify candidates. A total of 23 targets were selected by having kinematics consistent with the halo, defined here as follows: a proper motion greater than 0.01 
arcsec $yr^{-1}$ and a transverse
velocity greater than 200 km $s^{-1}$ 
\citep{Brown11}. An additional 41 targets were chosen from low-resolution spectra obtained by the SDSS-III collaboration for MARVELS target pre-selection in the \emph{Kepler} field. 
These spectra are similar to the spectra for 
the SEGUE and SEGUE2 surveys and were processed by the SEGUE stellar parameter pipeline, which has been shown to measure [Fe/H] with an uncertainty of 0.25 dex \citep{Lee08} and to successfully 
identify even metal-poor giants \citep{Lai09}, which are challenging to study at low resolution. Finally, 67 metal-poor candidates were selected on the basis of Washington photometry, 
which has the strongest metallicity sensitivity of any broadband 
system \citep{Canterna76,Geisler91}.  In combination with DDO51, these filters can reliably identify metal-poor giants.  We therefore selected a total of 128 candidates, of which 
27 were new objects added to the \emph{Kepler} LC sample in APOKASC GO proposal $\#40033$; there was some overlap in the lists of potential metal poor targets generated with the criteria above.  
The process for generating the metal-poor star 
candidate list and the yields from the various methods are described in Harding et al. (2014, in prep). Candidates in any of these three categories are referred to as ``Halo'' in the targeting flags.

\subsubsection {Rapid Rotators}

Rapidly rotating giants (counted as ``Rapid Rotators'' in Table 1)
 are relatively rare and may 
represent interesting stages of stellar evolution, such as recent mergers; we therefore screened the long-cadence sample for signatures of rotational modulation.  We found 162 targets
with such a signature in the sample.  

\subsubsection {Luminous Giants}

Intrinsically luminous giants 
(defined here as stars with $\log g$ $< 2$) are under-represented in the public giant sample relative to the field population because they were specifically selected against in the \emph{Kepelr}
planet transit survey design.  These stars are important targets for both stellar population and stellar physics studies 
because they extend the dynamic range in gravity for testing both asteroseismic scaling relationships and the seismic properties of giants.  More luminous giants in a magnitude-limited 
sample will be more distant from both Earth and the Galactic plane than less luminous ones, making the former more likely to be metal-poor. For our spectroscopic program we therefore selected all long-cadence
targets with KIC $\log g$ $< 1.6$, including 43 M giants and 122 other giants.  We also included all long-cadence giants with more than 8 quarters of data and $\log g$
between 1.6 and 2.2, adding 175 additional targets.  We also proposed new targets for long-cadence observations in a GO program that had 4 quarters of data (Q14-Q17) that were screened as
likely high-luminosity targets which met our magnitude and temperature cuts for APOGEE observation, using KIC properties as a basis. This dataset included
all stars with KIC $\log g$ between 0.1 and 1.1 and 20 stars randomly selected per 0.1 dex bin in $\log g$ between $\log g$ of 1.1 and 2.  This selection process added 253 new high-luminosity targets in total
to the APOGEE sample.  All candidates satisfying these criteria were counted as ``Luminous Giants'' in Table 1.

\subsubsection {Stars with Unusual Pulsation Properties}

Some stars, for poorly understood reasons, have unusually low $l=1$ mode amplitudes (\citealt{Mosser12b}; see \citealt{Garcia14} for a discussion), 
and spectroscopy of such targets is valuable.  We included 37 such targets.  There are 122 stars with good time coverage and 
unusually large $\Delta P$, and 15 such targets with unusually small $\Delta P$.  Included in our program are 90 stars whose $\Delta P$ values are intermediate between those expected for red 
clump and red giant branch stars, which is a possible signature of post-He flash stars \citep{Bildsten12}.  All of these targets
were prioritized for spectroscopic observation.

\subsection{Reference Sample Definition and APOKASC Sample Properties}

Our initial target list included all of the stars in the special categories described above.  We then defined a reference sample of stars with the 
highest possible quality of data: this included an accurate classification of evolutionary state and at most one quarter of data missing.  
We were left with 683 first-ascent giants; they comprise the bulk of the sample.  We randomly selected 150 red clump and 50 secondary clump 
stars from a larger pool (1684) of available candidates of comparable quality.  We supplemented this list with a secondary priority set (from the \emph{Kepler} GO program) 
of targets which otherwise fit our criteria but had less data available.
This added 39 first-ascent giants, 58 stars with detected rotation or unusual pulsation properties, 105 secondary red clump, and 227 red clump stars.

\begin{figure*}\plotone{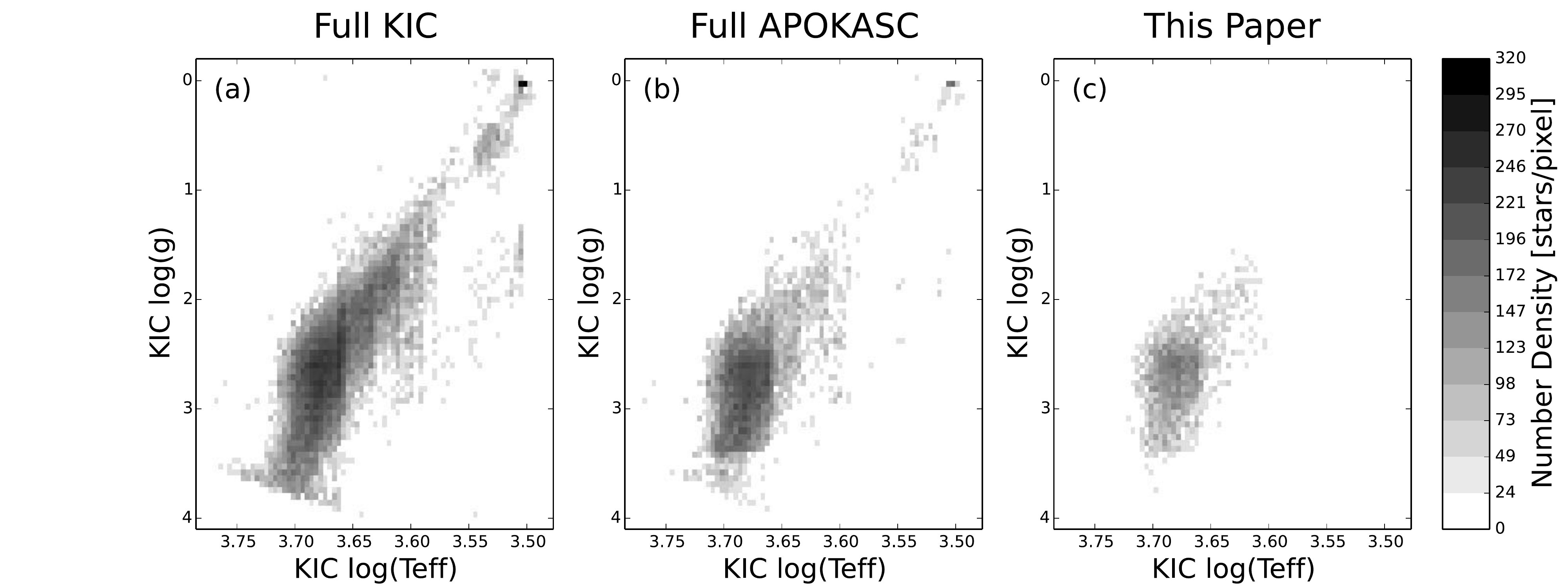}
  \caption{We compare stellar properties from the KIC of the full public and KASC giant sample (left), the full APOKASC sample (center), 
and the DR10 sample reported in this paper and summarized in Table 1 (right) in the HR Diagram.}
 \label{fig:sample}
 \end{figure*}

The remainder of the list was filled by public red giants (with or without data on their evolutionary state) and known red clump stars, identified as such by the diagnostics discussed in the previous section. 
The stars with ambigious evolutionary classification were divided into groups based on their 
asteroseismically-determined $\log g$. Public red giants with ambiguous evolutionary state measurements were considered possible red clump stars with $2.35<\log g<2.55$ and 
possible first-ascent red giant branch stars otherwise.  We allocated 
210 slots for red clump stars in 21 pointings.  If there were fewer than 10 known red clump stars in a given target APOGEE pointing, the remainder was filled with 
possible red clump stars. We then selected all of the possible first-ascent red giant branch stars. These were prioritized first by the number of quarters observed and then randomized among stars 
with the same length of observations.  Then, public giants that appeared as asteroseismic outliers were prioritized by H-band magnitude. This was 
followed by the rest of the known red clump stars, prioritized by brightness. Lastly, the public red giants identified as possible red clump stars were prioritized first by the number of quarters 
observed and then randomized among stars with the same length of observations.  We illustrate the net effect of our sample selection in Figure~\ref{fig:sample}.


Our final sample is described in Table 1.  Some stars are included in more than one category (for example,
luminous and halo); and there are 
1916 in total.  ``Gold'' refers to the targets used as calibrators for the spectroscopic surface 
gravity measurements in APOGEE.  The full set of targeting flags are included in our main catalog table (see Section 5).
 In this table we merge the two
 different labels of SEISMIC INTEREST and SEISMIC OUTLIER into one category, as both classes were targeted because of 
 unusual features in their measursed pulsation properties. Stars in the luminous giant category do not have 
 direct targeting flags in subsequent tables, but can be identified
 by their KIC surface gravity measurements.  The targeting flags for each of the other categories are indicated in
parens in the Category column.

There are two particularly important aspects of this sample worth noting.  First, a diverse
set of objects can be studied with this dataset; second, the overall sample is far from being a simple representation
of the underlying stellar population.  Basic statistics, such as the ratio of core He-burning to shell H-burning targets,
are biased relative to the underlying populations. Not all stars targeted as members of a class are true members.
As a concrete example, only a portion of prospective halo giants satisfied our high-resolution abundance criteria.
We therefore urge that care be employed when using our data for stellar population studies.  The full DR12 sample is both more comprehensive and, 
arguably, less customized to individual projects; we defer a full discussion of
the impact on population studies to the release of the second dataset.
For completeness, there are three other samples selected in this process and observed in APOGEE that will be released separately because they are 
either outside of the \emph{Kepler} fields or do not involve red giants.  We observed all \emph{Kepler} Objects of Interest, or KOIs, that fit our magnitude and color cuts, 
all dwarfs within our magnitude range that had $T_{\rm eff} < 5500$ K, and selected targets observed by the CoRoT satellite \citep{Mosser10}.  
A discussion of their properties is beyond the scope of this paper.

\begin{deluxetable}{lr}
\tablecaption{Breakdown of APOKASC Targets in DR10 }
 \tablehead{ \colhead{Category} & \colhead{(Non-Unique) Number}}
 \startdata
 Gold (GOLD)                         &   286     \\
 KASC (KASC)                         &   678     \\
 Halo (HALO)                         &   40      \\ 
 Luminous Giant (LUMINOUS)           &   115     \\ 
 Cluster        (CLUSTER)            &   43      \\
 Seismically Interesting or Outlier      &   221     \\
 RC (Seismically Classified) (RC)   &   204     \\ 
 RGB (Seismically Classified) (RGB) &   68      \\
 Rapid Rotator (ROTATOR)            &   17      \\
 Total                              &   1916    \\
 \enddata\label{table:Results}
 \end{deluxetable}

\section{Spectroscopic Properties}

The APOGEE Stellar Parameters and Chemical Abundances Pipeline (ASPCAP) 
was employed to infer six atmospheric parameters from the 
observed spectra: effective temperature ($T_{\rm eff}$), metallicity ([M/H]), 
surface gravity ($\log~g$), carbon ([C/M]), nitrogen ([N/M]), and $\alpha$ ([$\alpha$/M])
abundance ratios.  Automated pipeline analysis is powerful, but there is always the possibility of systematic errors
in the derived parameters.  \citet{Meszaros13} therefore checked the spectroscopic measurements of surface gravity,
effective temperature, and metallicity against literature values using 559 stars in 
20 open and globular clusters and derived 
corrections to place our results on the same system as these measurements.  
A limited subset of asteroseismic surface gravities were also used to calibrate results for higher metallicity stars.
The net result
are two spectroscopic scales: the ``raw'' pipeline values and the ``corrected'' ones that were 
a result of the calibration procedure. 
Details of the spectroscopic pipeline, its calibration, and 
the exact equations for the correction terms are presented by \citet{Meszaros13}.


The APOGEE spectroscopic 
$T_{\rm eff}$ and metallicity values were used as inputs for deriving the asteroseismic masses, radii,
surface gravities, and mean densities as described in Section 4.  We derived and present results for both the raw
and the calibrated spectroscopic scales.  In this section of the paper we use the
APOKASC sample to provide two new checks on random and systematic uncertainties in the spectroscopic surface gravities
and effective temperatures from both asteroseismology and photometry.  We compare spectroscopic and asteroseismic 
surface gravities in our full dataset, which is a much larger sample than that used in the \citet{Meszaros13} work.  
The \emph{Kepler} fields are relatively low in extinction, and a complete set of \emph{griz} and \emph{JHK} photometry is
available for our targets (from the KIC and 2MASS respectively).  Our comparison here of photometric and spectroscopic $T_{\rm eff}$ measurements,
therefore provides a good external check on the KIC extinction map used to derive them and on the absolute
spectroscopic temperature scale.  Finally, the ASPCAP calibration procedure derived independent corrections
to the three major spectroscopic parameters considered here.  In principle, one could instead have 
imposed an external prior on one or more of them, and then searched for a refined spectroscopic solution.
Because we have precise asteroseismic surface gravities, we quantify here how 
our metallicities and temperatures would have been impacted if 
we had adopted them as a prior rather than independently calibrating all three.
A detailed comparison of the full sample results with the KIC and optical spectroscopy, which were not
used to calibrate our measurements, is presented in Section 5.


\subsection{The $T_{\rm eff}$ Calibration}

The ASPCAP spectroscopic effective temperatures were compared with photometric ones using 
calibrations by \citet{Gonzalez09}, hereafter GHB09, using 2MASS $J-K_{s}$ colors \citep{Skrutskie06}.
\citet{Meszaros13} found systematic differences in the range of 100 K to 200 K 
between the raw ASPCAP and those derived using the GHB09 scale and literature extinction estimates
for star clusters. 
ASPCAP $T_{\rm eff}$ values were found to be consistent with literature values from optical
spectroscopy, in mild tension with the photometric values.  \citet{Meszaros13} recommended 
calibrating the ASPCAP $T_{\rm eff}$ to conform with the photometric scale, because it
is closer to the fundamental definition of the effective temperature than the spectroscopic studies; however,
the purely spectroscopic scale is defensible, and the difference between the two serves as a measure
of plausible system zero-point shifts. 
The ASPCAP temperatures were corrected between 
3500 K and 5500 K using an equation derived from the comparison with the GHB09 
scale. The GHB09
calibration was chosen as a calibrator because it is explicitly designed for red giants. It
is within $30-40$ K of the recent dwarf-only Infrared Flux Method (IRFM) temperature scale of
\citet{Casagrande10}.
Photometry provides precise relative temperatures for cluster members, and the RMS differences between the
ASPCAP and photometric temperatures for such stars were used by \citet{Meszaros13} as a measure of random
spectroscopic temperature uncertainties.  Random uncertainties ranged from $\sim 200$ K for metal-poor stars to
$\sim 100$ K for metal-rich ones.

The majority of the cluster stars used
for the calibration are metal poor, while our sample has a mean metallicity close to solar.  It is therefore
possible that there could be metallicity trends in the temperature differences, a topic that we explore below
when we compare photometric and spectroscopic temperatures in the \emph{Kepler} fields.

\subsection{The Metallicity Calibration}

The [M/H] dimension in ASPCAP was constructed by varying the solar-scaled abundances of all elements except
C, N, and the $\alpha$-capture elements O, Mg, Ca, Si, Ti.  Therefore, the best-fix [M/H] represents a
line-weighted fit to the iron peak and light odd-Z elements.  In practice, as discussed below, [M/H] correlates
well with [Fe/H], and we will treat them as being functionally equivalent in this paper.  The ASPCAP
metallicity was compared with individual values from high-resolution 
observations from the literature, and with average cluster values. The derived 
metallicities from ASPCAP are close to literature values around solar metallicity. 
The difference in cluster averages between ASPCAP and literature becomes larger than 
0.1 dex only below [M/H] $= -1$, and this discrepancy increases with decreasing metallicity, 
reaching $0.2-0.3$ dex around [M/H]$= -2$ and lower. An offset of comparable magnitude ($\sim 0.1$ dex) 
was found above [M/H] $= +0.1$. \citet{Meszaros13} therefore derived a calibration to bring the raw
metallicities into agreement with the literature cluster averages. 

The metallicity uncertainty was derived from the standard deviation of individual cluster member metallicities around the cluster averages. 
This scatter can be as high as 0.14~dex for the lowest metallicity globular clusters, 
but it improves significantly (similarly to $T_{\rm eff}$) for high metallicities. 
For open clusters around solar metallicity, the largest scatter is only 0.07~dex.
We compare our metallicities with those derived from optical spectroscopy and from the KIC in
Section 5. 

\subsection{The Surface Gravity Calibration}

Surface gravities can be estimated from isochrones for red giants in star clusters if the distances, extinctions, and ages of the systems are known.
\citet{Meszaros13} found significant zero-point offsets between the raw spectroscopic values and those derived from cluster isochrones, motivating
an empirical correction.  The cluster-based surface gravity calibration  was supplemented with a preliminary APOKASC sample of asteroseismic gravities.  
The \emph{Kepler} targets are concentrated around solar metallicity, while the cluster sample is predominantly composed of metal-poor systems.
We therefore adopted a hybrid empirical calibration that was solely a function of metallicity and solely based on the gold standard asteroseismic
surface gravities for stars with  [Fe/H] $> -0.5$.

This gold standard sample had to be defined prior to the full analysis, and we briefly describe how this sample was assembled and analyzed below.
The candidates were selected to be those with the most complete time coverage; see \citet{Hekker12} for a
discussion of the criteria.  We then computed the mean asteroseismic parameters using the \citet{Hekker10} methodology.
We adopted effective temperatures based on the \emph{griz} SDSS filters \citep{Fukugita96} from \citet{Pinsonneault12}, using the KIC extinction map.  
Grid modeling was performed using BaSTI models and adopting the KIC metallicities.  We added 0.007 dex in quadrature to the formal
uncertainties to account for systematic errors; see \citet{Hekker13} for a discussion.  A total of 286 stars from this list were observed in DR10
and used as calibrators; see \citet{Meszaros13} for a more complete discussion.  We present the asteroseismic properties used for this
calibrating sample in Table 2.  We assess the validity of this approach using the full asteroseismic surface gravity from our data below.
Table 2 lists the surface gravities derived for the gold standard candidates.  These were used to calibrate the ASPCAP spectroscopic surface gravities, and are included so that the results of
\citet{Meszaros13} can be replicated.  The first column contains the KIC ID. The second is the effective temperature inferred from SDSS photometry in \citet{Pinsonneault12}. The
third column is the KIC metallicity, and the fourth is the logarithm (in cgs units) of the surface gravity returned from the OCT pipeline and its uncertainty.
We stress that the gravities presented later in the paper supercede these values.

\begin{deluxetable}{lrrr}
\tablecaption{Gold Standard Surface Gravities}
\tablehead{\colhead{KIC ID} & \colhead{$T_{\rm eff}$} & \colhead{[Fe/H]} & \colhead{$\log g$}\\
\colhead{} & \colhead{(K)} & \colhead{} & \colhead{(cgs)}}
\startdata
1161618 & 4907 & -0.112 & 2.426	$\pm$ 0.010 \\
1432587 & 4693 & -0.022 & 1.661	$\pm$ 0.024 \\
1433593 & 5013 & -0.141 & 2.741	$\pm$ 0.014 \\
1433730	& 4829 & -0.099 & 2.504	$\pm$ 0.013 \\
1435573	& 4943 & -0.113 & 2.324	$\pm$ 0.012 \\\enddata \label{table:Goldlogg}
\end{deluxetable}

\subsection{Tests of the APOKASC Temperature and Surface Gravity Calibrations}

\begin{figure*}\plottwo{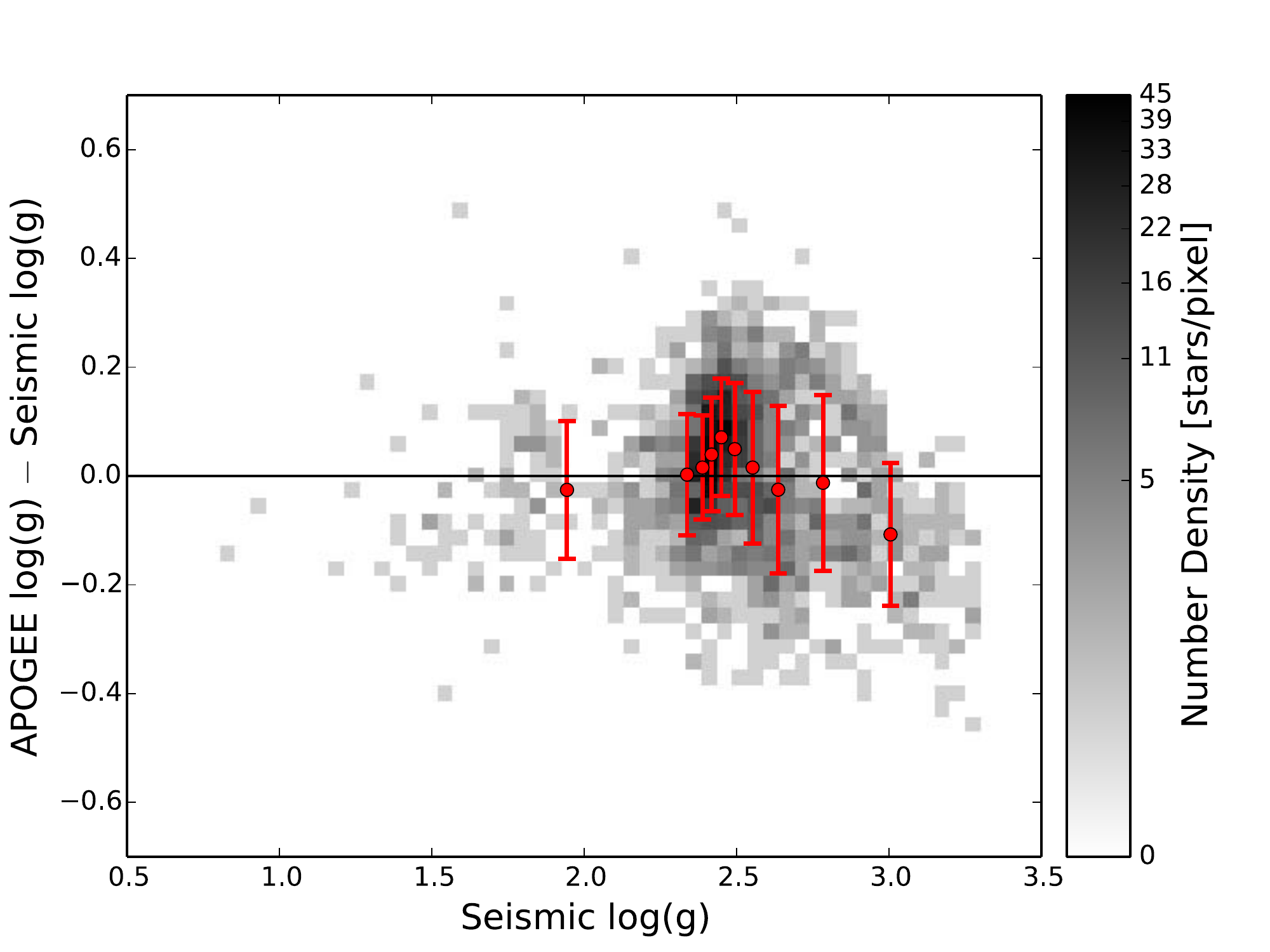}{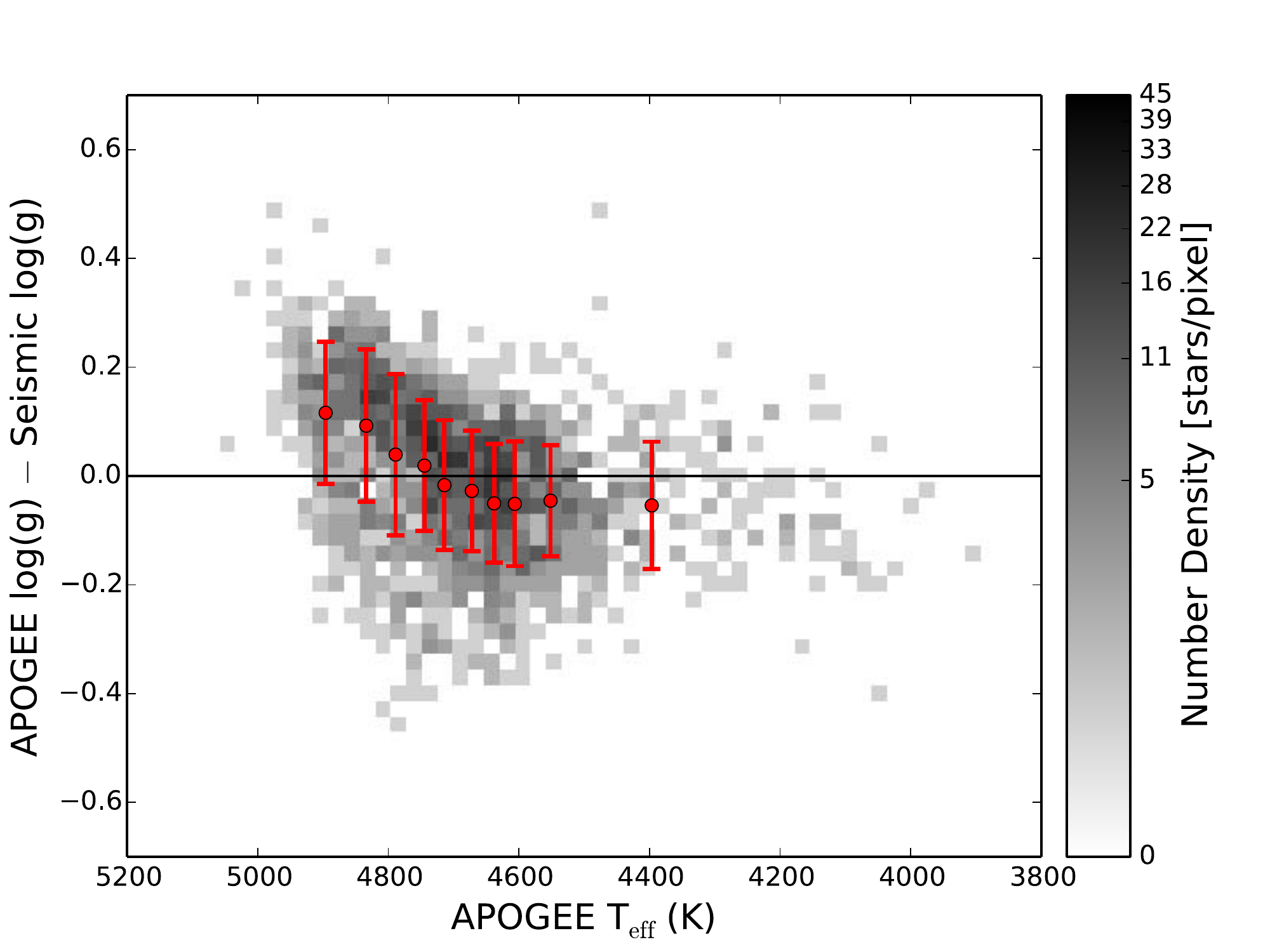}
  \caption{Logarithmic difference between the corrected spectroscopic and asteroseismic surface gravity $\log g$
as a function of asteroseismic $\log g$ (left) and spectroscopic $T_{\rm eff}$ (right) for our full sample.  
The points with error bars are the means and standard deviations of the data in 10 ranked cohorts of $\log g$.  
The data was divided into 60 bins in $\log g$ (left), $T_{\rm eff}$ (right) and delta $\log g$ (both), covering the $\log g$ range of 0.5 to 3.5,
temperature range 3800 K to 5200 K, and gravity difference range $-0.7$ to $+0.7$ respectively.
The logarithmic gray scale coding (specified on the right) indicates the number of targets with those properties in the relevant bin.}
 \label{fig:deltagvsg}
 \end{figure*}

In the APOKASC sample we have access to 
extremely precise asteroseismic surface gravities \citep{Hekker13}, which we adopted for the catalog in preference to 
the spectroscopic solutions. However, the calibration procedure included only a smaller subset of the data, the gold standard asteroseismic sample, as a reference.
Our mean calibrated spectroscopic and asteroseismic $\log g$ values for the full sample are close, with an average offset of 0.005 dex and a dispersion of 0.15 dex.  
We therefore conclude that our calibration based on the limited preliminary dataset yielded reasonable results on average for the full sample (in the sense that
the typical differences between our calibrators and spectroscopic values were similar to the same differences for stars not used as calibrators).  However, there are 
interesting underlying trends in the differences between asteroseismic and spectroscopic $\log g$, present in both the gold and full samples, which are illustrated
as a function of $\log g$ and Teff in Figure~\ref{fig:deltagvsg}.


 \begin{figure*}\plottwo{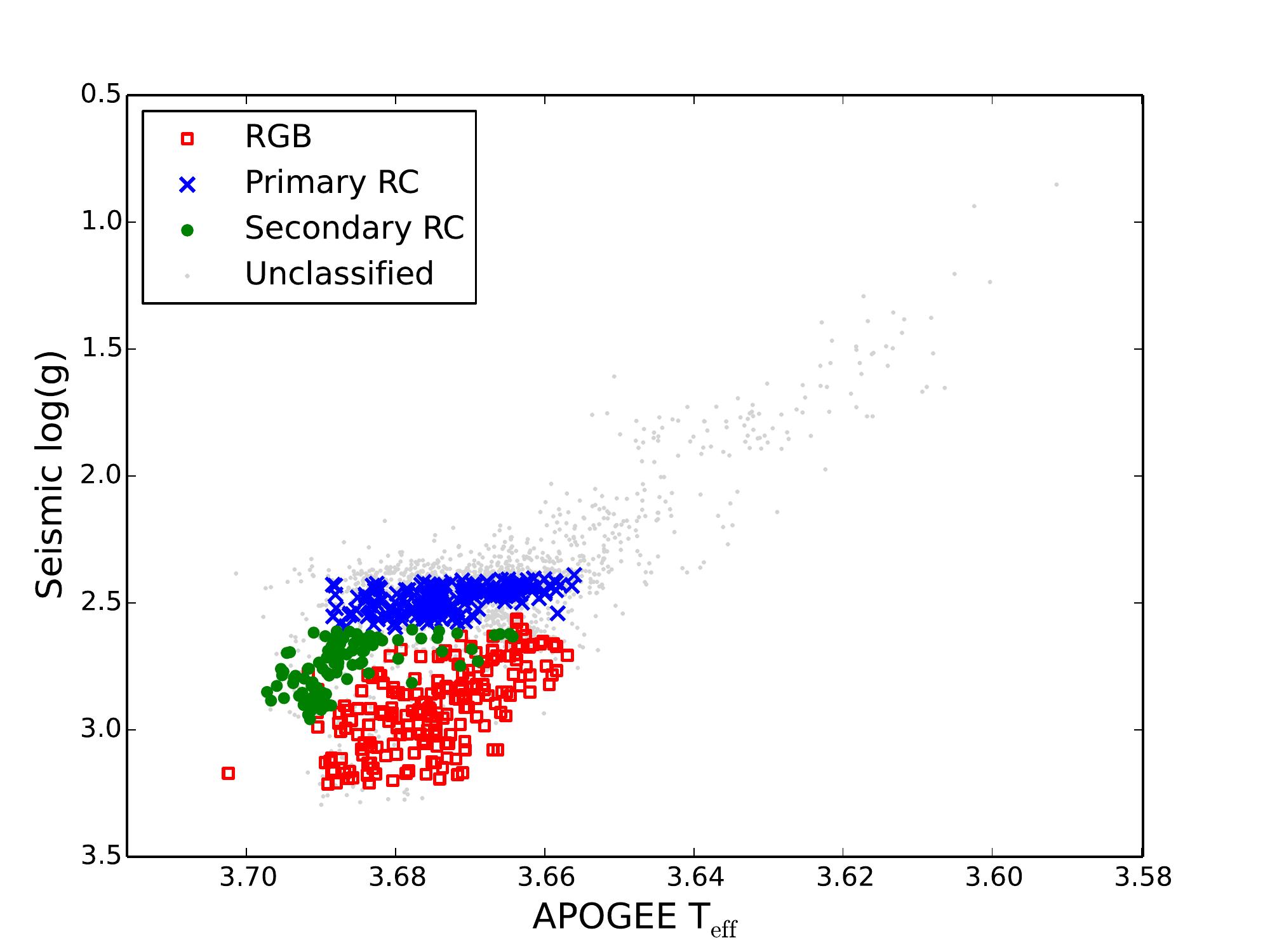}{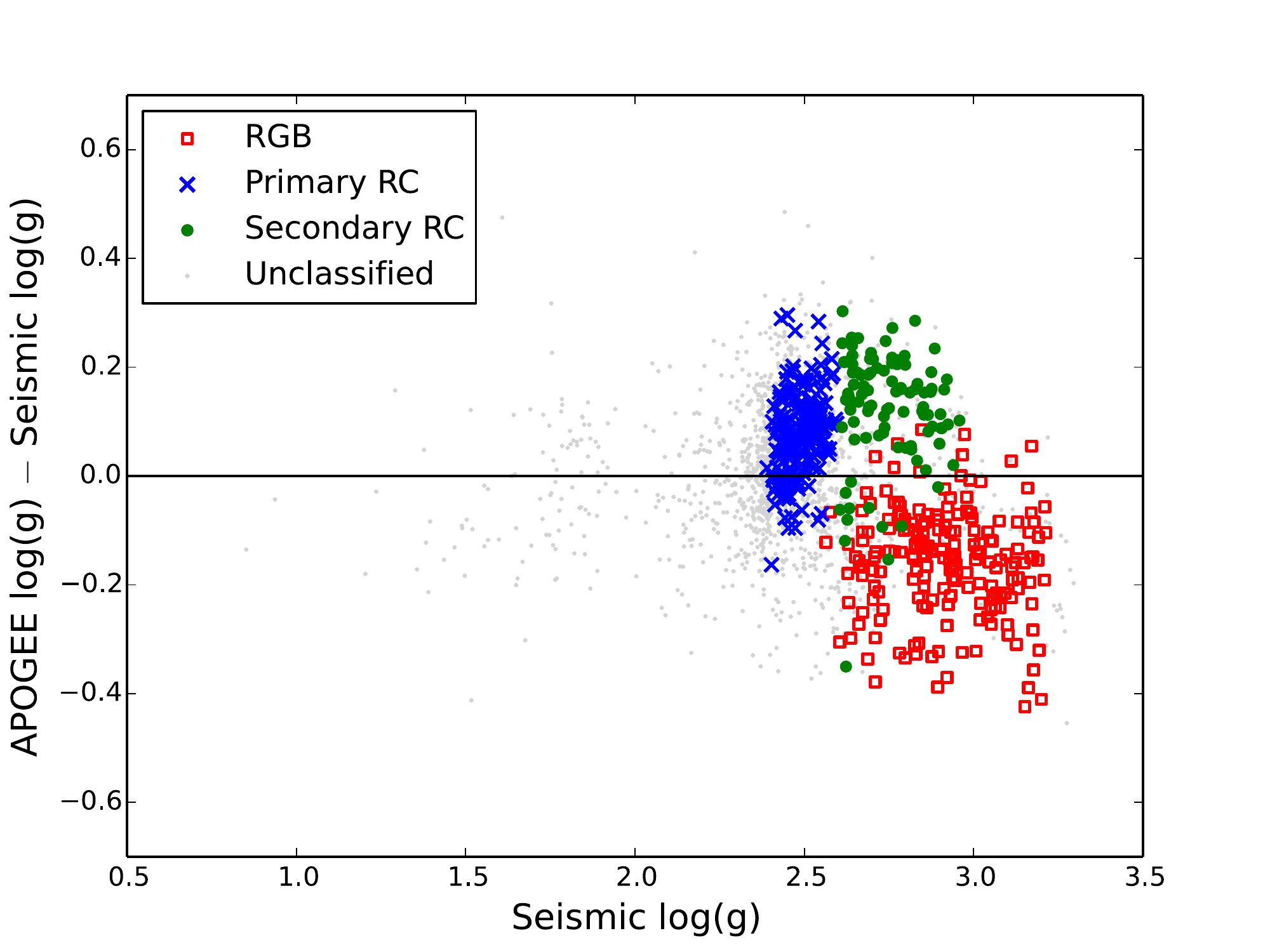}
  \caption{Stars identified asteroseismically as secondary RC (green), RC (blue), or first-ascent RGB (red) 
are compared in the HR Diagram on the left.  Gray dots represent stars without explicit classification.
On the right, the difference between asteroseismic surface gravity
and spectroscopic surface gravity is plotted for stars in these evolutionary states as a function of asteroseismic $\log g$.  A clear pattern
in the differences is visible.}
 \label{fig:evstatecompare}
 \end{figure*}


We can obtain further insight into the origin of these differences by adding information on evolutionary state.  When we do so, a clear division in mean difference emerges
between core He-burning, or red clump, stars and first-ascent red giant branch stars (see Figure~\ref{fig:evstatecompare}).  Some of these differences can be traced
to the temperature and surface gravity trends illustrated above, but the differences persist even between members of the same star cluster.  We are currently investigating
the origin of the gravity offsets between red clump and red giant branch stars.  Small differential offsets (at the $ 5\%$ level)  between asteroseismic radius estimates for red clump and red giant
branch stars were found in NGC 6791 \citep{Miglio12} and traced to differences in the sound crossing time at fixed large frequency spacing.  However, the impact of the Miglio corrections on the relative
radii are too small to explain the observed surface gravity discrepancy, 
and the differential offset in surface gravity is likely to be at the 0.05 dex level or smaller because of correlations between asteroseismic masses and radii.  An offset between the structures of model atmospheres in stars
with similar HR diagram position but with different evolutionary states (and thus differences in mass, helium, or CNO) is in principle possible; however, such effects are expected to be small.
We conclude that this offset is an interesting clue that may shed light on both model atmospheres and asteroseismology, but that the mean values of the corrected asteroseismic and spectroscopic gravities are in good agreement.

\begin{figure*}\plottwo{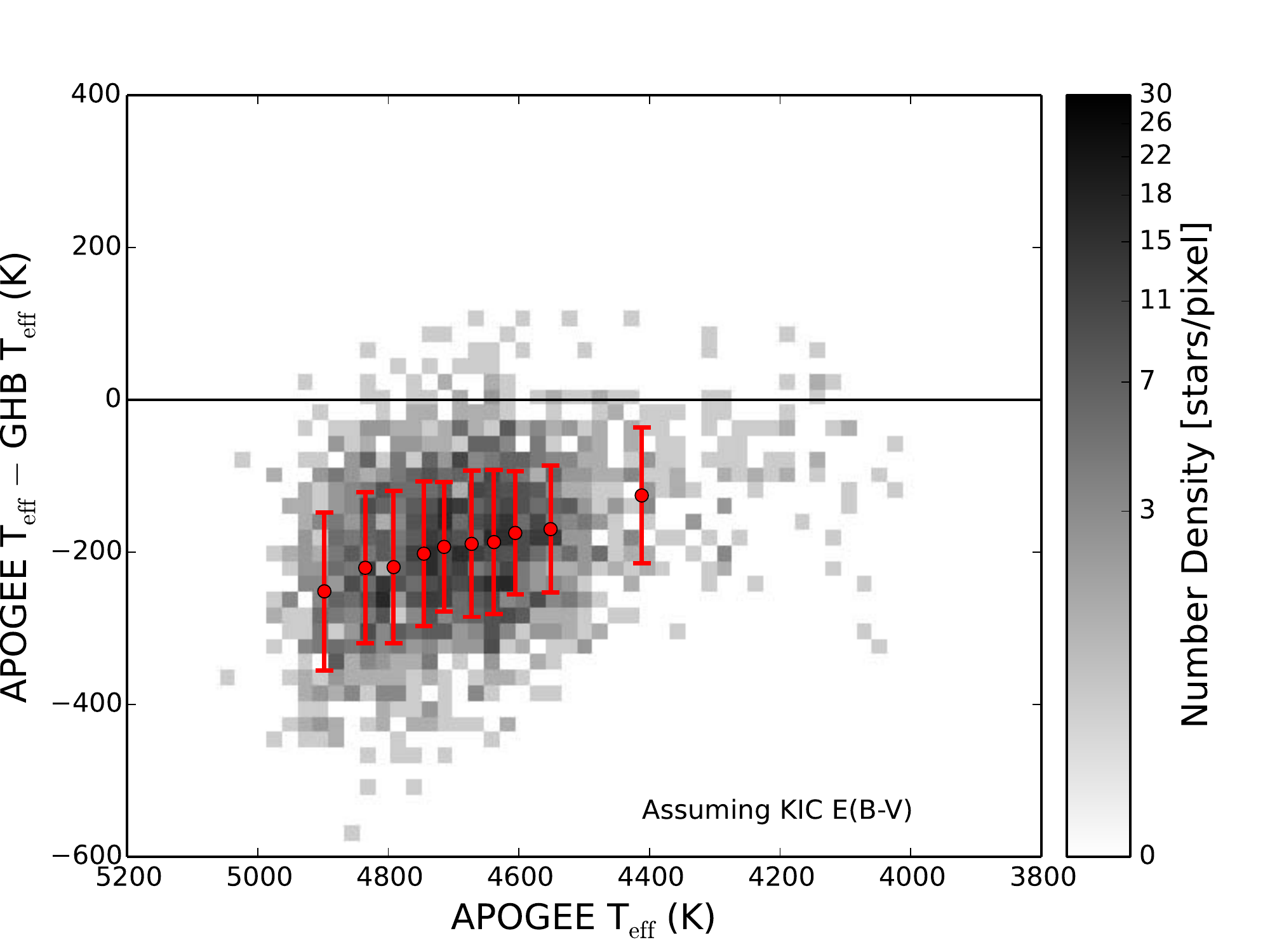}{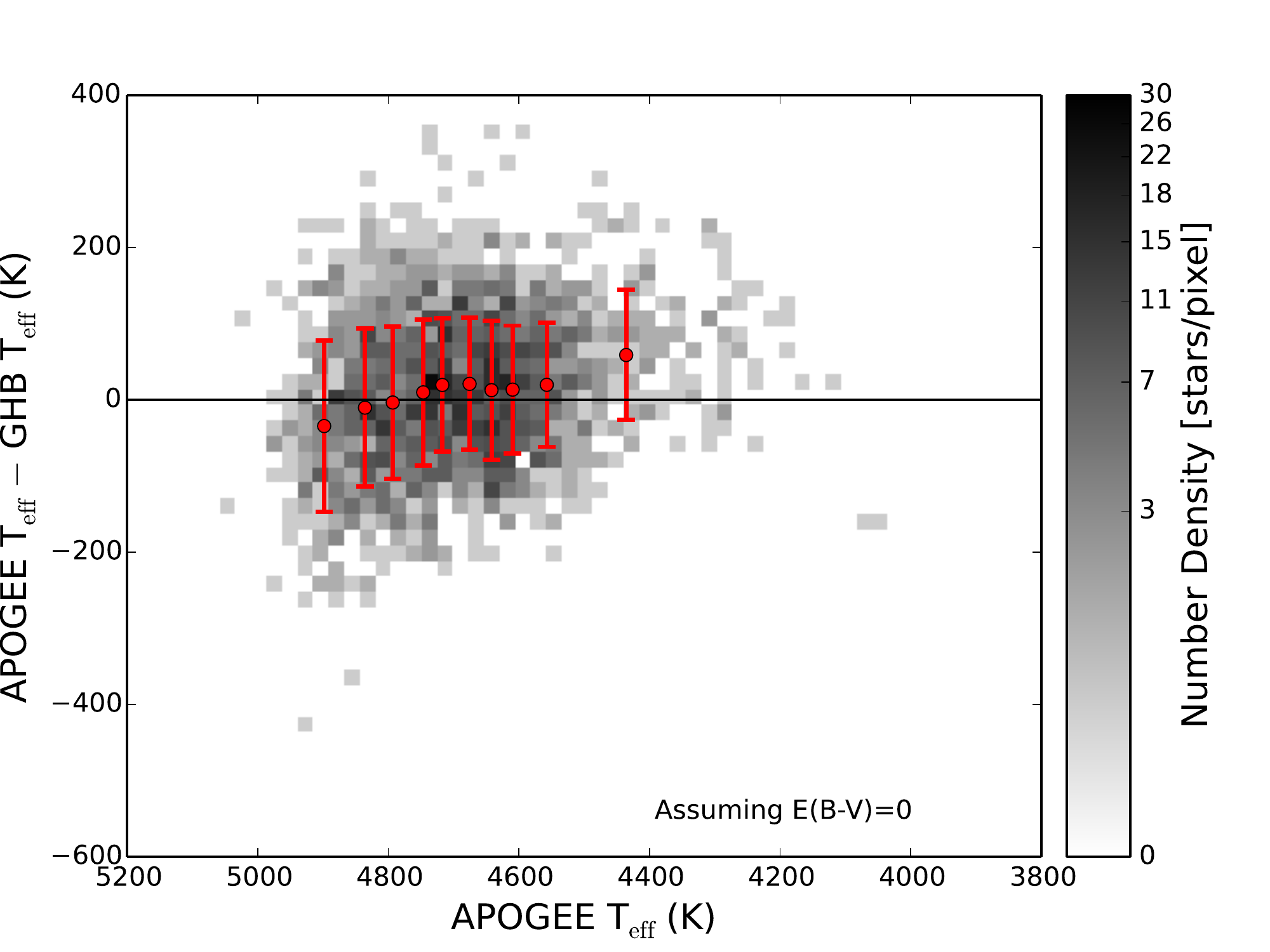}
  \caption{Differences in effective temperature $T_{\rm eff}$
as a function of spectroscopic $T_{\rm eff}$ for our full sample.  Photometric temperatures were computed
using the KIC extinction map (left) and for zero extinction (right), and with the spectroscopic [M/H], 2MASS JK colors, and the GHB09 color-temperature
relationship.  The points with error bars
are the means and standard deviations of the data in 10 ranked cohorts of $T_{\rm eff}$.  
The data was divided into 60 bins in delta $T_{\rm eff}$  and $T_{\rm eff}$ covering the range $-600$ K to $+400$ K and 3800 K to 5200 K respectively,
and the logarithmic color coding (specified on the right) indicates the number of targets with those properties in the relevant bin.}
 \label{fig:deltatkic}
 \end{figure*}

We can also check on the internal consistency of the spectroscopic temperature scale by comparing our spectroscopic effective temperatures with those that
we would have derived using the KIC extinction map and the GHB09 IRFM color-temperature relationship employed in the global spectroscopic calibration.
Our results using the KIC extinction map are compared with those in the zero-extinction limit in Figure~\ref{fig:deltatkic}.  The dispersion is reasonable at $80$ K, but there is a significant
zero-point offset of $-193$ K in the former case.  The bulk of the calibrating sample was in metal-poor globular cluster stars, so this feature could reflect a metallicity-dependent offset
in the temperature scale; the APOKASC sample is predominantly close to solar abundance.  Another possibility is an error in the adopted extinction corrections; as shown above, a
zero-extinction case has an average offset of $+11$ K.  We view this as an unrealistic limit, but the difference between the two certainly highlights the need for an independent re-assessment of the KIC extinction map.
Fortunately, a more extensive multi-wavelength dataset, especially at longer
wavelengths, has been developed since the time the KIC was constructed. \citet{Casagrande14} used new Stroemgren filter data in a stripe within the \emph{Kepler} fields, along with an extensive set of
literature photometry, to derive systematically smaller extinction values than those in the KIC. Adopting their extinction map  would imply a smaller (but real) temperature offset.  
 \citet{Zasowski13} used 2MASS, IRAC, and WISE data to infer extinctions when developing the APOGEE
target list, using the RJCE method \citep{Majewski11}; work on a related approach for \emph{Kepler} targets is in progress.  \citet{Rodrigues14} employ a related method of SED fitting and 
confirm a lower extinction estimate than that obtained from the KIC alone.  
Adopting the latter extinction values implies a $T_{\rm eff}$ difference of 74 K between the corrected ASPCAP and GHB IRFM scale for solar abundance stars.  

\begin{figure*}\plotone{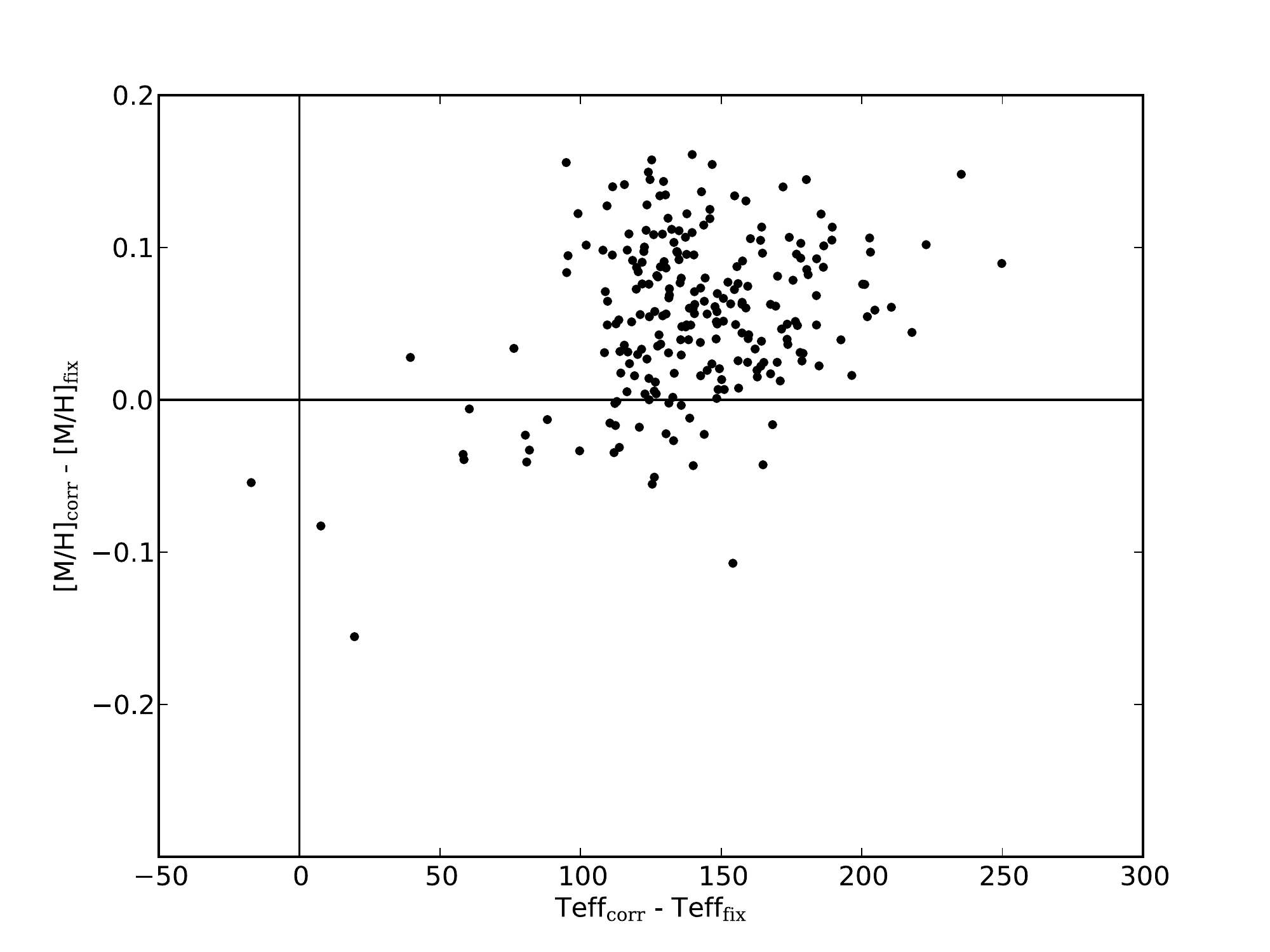}
  \caption{Differences in effective temperature and metallicity between the ASPCAP spectroscopic solution for a fixed asteroseismic surface gravity (subscript fix) 
and the corrected ASPCAP results (subscript corr).  The spectroscopic parameters obtained with an 
asteroseismic prior are systematically offset to lower metallicity and effective temperature relative to those calibrated against independent diagnostics.}
 \label{fig:corr_fix}
 \end{figure*}

Finally, we have derived independent calibrations of temperature, metallicity, and surface gravity, applied after the ASPCAP parameter solution was obtained.  
An alternate method that has been successfully used for 
dwarfs in the \emph{Kepler} fields is to to search for the best solution adopting the asteroseismic surface gravity as a prior \citep{Chaplin14}.  
This approach is similar in philosophy to using isochrone fits to surface gravities in star cluster dwarfs rather than searching for less precise spectroscopic values
for warm dwarfs.  As a test of adopting this approach, we used the ASPCAP pipeline for all parameters except surface gravity, for which we supplied the asteroseismic values.  
The chi-squared metric for the best fit was visibly degraded, as expected.  However, the resulting metallicities and effective temperatures were also offset 
from the values obtained from our independent calibration checks.  This effect is illustrated in Figure ~\ref{fig:corr_fix} where we compare the values that we would 
have obtained with an asteroseismic gravity prior with the actual calibrated values.  The sense of the difference in temperature is expected from Boltzmann-Saha 
balance considerations, and the corrected values using this approach are actually in worse agreement than the raw ones when compared with independent measurements.  
We therefore conclude that the \citet{Meszaros13} approach of independent calibrations for each of the spectroscopic parameters is more accurate for our purposes than 
adopting an asteroseismic surface gravity prior.

In summary, our raw spectroscopic parameters have been derived using a homogeneous analysis method.  Comparisons with independent measurements motivated us to define modest 
correction terms for metallicity and effective temperature and more substantial ones for the spectroscopic surface gravities.  At the metal-rich end, the spectroscopic surface gravities
were tied to an asteroseismic reference scale using a limited sample of gold standard targets.  With the full DR10 sample of spectroscopic and asteroseismic data we
revisited the spectroscopic surface gravity calibration with a much larger sample of asteroseismic surface gravities.  The larger asteroseismic sample is in good mean agreement with 
the \citet{Meszaros13} calibration, but there are modest (but real) systematic offsets at the $\sim 0.1$ dex level between the asteroseismic and spectroscopic scales as functions of effective temperature, gravity,
and evolutionary state.  Further work is needed to identify the origin of these effects (in terms of systematics in either the asteroseismic or spectroscopic surface gravities).

We also compared our spectroscopic effective temperatures 
for \emph{Kepler} field red giants against $T_{\rm eff}$  derived from the same photometric temperature calibration that was used in star clusters.  
There is also an offset of 193 K between the spectroscopic $T_{\rm eff}$ scale and the IRFM photometric $T_{\rm eff}$ scale if the KIC extinction map is adopted; however,
there is independent evidence that the KIC extinctions are overestimated, so this result should be treated as an upper bound on systematic $T_{\rm eff}$ errors for our sample.  
With the extinction map of \citet{Rodrigues14} we can quantify the zero-point shift more precisely (and it is at the 74 K level).  If this offset is confirmed,
it implies a metallicity-dependent temperature correction that was not captured in the original calibration.  We view these temperature and gravity comparisons as 
fair indicators of potential systematic uncertainties in these properties.  Neither of these comparisons directly address the accuracy and precision of our metallicity estimates, 
which we discuss in Section 5 when comparing our final catalog values with the KIC and optical spectroscopy.

\section {Asteroseismic Properties}

Our goal for the asteroseismic sample was to provide masses, radii, and surface gravities for
all of our red giants.  This is a complex task with multiple stages, and we discuss our
methodology below; we begin with a brief summary.  

We processed and corrected the raw light curves to extract
the oscillation frequencies.  We then identified the modes, and distilled the information
from the observed oscillation pattern
down to two global asteroseismic parameters: the frequency of maximum power and the average large 
frequency spacing, whose interpretation is discussed below.  These parameters, combined with
the effective temperature, could in principle be used to solve for the asteroseismic masses and
radii with the aid of scaling relations.  We took the additional step of combining information
from stellar models with the global asteroseismic parameters, and used this grid modeling effort to both identify outliers and to
refine our estimates.  In all cases we employed multiple methods and made an attempt to assess
random and systematic uncertainties.  We release asteroseismically determined stellar parameter estimates for two distinct
spectroscopic inputs, corresponding to the raw and corrected scales discussed above.

 \subsection{Light Curve Data Analysis}
 \label{sec:data}

Preparation of \emph{Kepler} long-cadence data \citep{jenkins10}
for asteroseismic analysis was handled in three cohorts.  
We used PDC-SAP (Pre-search Data Conditioning Simple Aperture
Photometry) light cuves \citep{Smith12,Stumpe12} for a cohort of 2067 field
stars.  These data were prepared for asteroseismic analysis in
the manner described by \citet{garcia11}. \emph{Kepler} data for quarters
Q0 through Q8 were used for these targets.  A second cohort of
657 field stars had pixel-level data available from \emph{Kepler}
observing quarters Q0 through Q12 inclusive. Aperture photometry 
was performed on these data (Mathur, Bloemen et al. in prep), producing 
light curves that were more stable at longer time scales than the
PDC-MAP data.  This improvement was at the expense of having slightly elevated
high-frequency noise.  Finally, raw Q0 through Q11 data were prepared for
34 stars in the open cluster NGC\,6819, again using the procedures
described in \citet{garcia11} (see also \citet{Stello11b}). A
total of 581 stars had data available in both the pixel and the public
sets described above (a point we return to briefly at the end of
Section~\ref{sec:cat}).

In total, five data analysis methods - \citet{Huber09},
\citet{Hekker10}, \citet{Kallinger10}, \citet{mathur10}, and
\citet{Mosser11a} -  were used to extract independent estimates of two
global asteroseismic parameters from the frequency-power spectrum of
the lightcurves. Some of the analysis methods were not
applied to every dataset. One parameter was the average large
frequency separation, $\Delta \nu$ , the mean spacing between consecutive
overtones of the same angular degree, $l$.  The average large
separation scales to very good approximation as $\rho^{1/2}$, $\rho
\propto M/R^3$ being the mean density of a star having mass $M$ and
surface radius $R$ (e.g., see \citealt{Tassoul80}; \citealt{Ulrich86};
\citealt{jcd93}). The dependence of $\Delta \nu$ on the mean
stellar density may be used as a scaling relation normalized by solar
properties and parameters, i.e.,
  \begin{equation} 
  \frac{\Delta \nu }{\Delta \nu _{\odot }} \simeq 
       \sqrt{\frac{M/M_{\odot }}{(R/R_{\odot })^{3}}}.
  \label{eq:dnu}
  \end{equation}
The second global parameter is $\nu_{\rm max}$, the frequency of maximum
oscillation power. It has been shown to scale to good approximation as
$gT_{\rm eff}^{-1/2}$ \citep{Brown91, Kjeldsen95, Chaplin08, Stello09b, Belkacem11}, 
where $g$ is the surface
gravity and $T_{\rm eff}$ is the effective temperature of the
star. The following scaling relation may therefore be adopted:
  \begin{equation} 
  {{\nu _{\rm max}}\over {\nu _{\rm max,\odot }}}\simeq 
  {{M/M_{\odot }}\over {(R/R_{\odot })^2\sqrt{(T_{\rm eff}/T_{{\rm eff},\odot })}}}.
  \label{eq:numax}
  \end{equation}

The completeness of the results (i.e., the fraction of stars with returned
estimates) varied, since some pipelines are better suited to
analysing different ranges in \numax.

We selected one data analysis method, OCT\citep{Hekker10}, 
to provide the catalog global asteroseismic parameters of the stars
in all three cohorts. This selection was based on the returned
$\nu_{\rm max}$ values. The Hekker et al. method had the highest
completeness fraction, and results that were consistent with those
given by the other pipelines.  This approach ensured we obtained a
homogenous set of global asteroseismic parameters, which were then
used to estimate the fundamental stellar properties (see
Section~\ref{sec:griddling} below).

For outlier rejection we selected a reference
method for each cohort, this being the one whose $\nu_{\rm max}$ estimates
lay closest to the median over all stars in that cohort (Mosser et
al. for the pixel and public cohorts; and Kallinger et al. for the
cluster cohort). If the Hekker et al. $\nu_{\rm max}$ differed from the
reference $\nu_{\rm max}$ by more than 10\,\%, we rejected the asteroseismic
parameters for that star. This procedure removed 28 stars from the pixel cohort,
167 stars from the public cohort, and 1 star from the cluster cohort.

Uncertainties on the final $\Delta \nu$ and $\nu_{\rm max}$ of each star were obtained
by adding, in quadrature, the formal uncertainty returned by the
Hekker et al. method to the standard deviation of the values
returned by all methods. We also allowed for known systematic errors
in Equation~\ref{eq:dnu} (e.g., see \citealt{White11} and \citealt{Miglio12}), 
by including an additional systematic contribution of
1.5\,\% (also added in quadrature). Because $\Delta \nu$ is usually
determined more precisely than $\nu_{\rm max}$, we also added the same
systematic contribution to the $\nu_{\rm max}$ uncertainties. This is
essentially the approach adopted by \citet{Huber13} in their
analysis of asteroseismic \emph{Kepler} Objects of Interest.

 \subsection{Grid-based Modeling}
 \label{sec:griddling}

For each star we used the two global asteroseismic parameters,
$\Delta \nu$ and $\nu_{\rm max}$, together with the estimates of effective temperature
$T_{\rm eff}$ and metallicity [Fe/H], as input to ``grid-based''
estimation of the fundamental stellar properties. This approach matches the set of
observables to theoretical sets calculated for each model in an
evolutionary grid of tracks or isochrones. The fundamental properties
of the models (i.e., $R$, $M$ and $T_{\rm eff}$) were used as inputs
to the scaling relations (Equations~\ref{eq:dnu} and~\ref{eq:numax})
to calculate theoretical values of $\Delta \nu$ and $\nu_{\rm max}$ for matching
with the observations.

Every pipeline adopted solar values
$\Delta\nu_\odot=135.03\,\mu$Hz and $\nu_{\rm max,\odot}=3140\,\mu$Hz,
which are the solar values returned by the pipeline we selected to
return final values on our sample 
\citep{Hekker10}. The uncertainties in $\Delta\nu_\odot$
($0.1\,\rm \mu Hz$) and $\nu_{\rm max,\odot}$ ($30\,\rm \mu Hz$) were
accounted for by increasing the uncertainties in the $\Delta \nu$ and
$\nu_{\rm max}$ data of each star, using simple error propagation. Further
details on grid modeling using asteroseismic data may be found in,
for example, \citet{Stello09}, \citet{Basu10,Basu12}, \citet{Gai11} and \citet{Chaplin14}. 

We adopted a grid-based analysis that coupled six pipeline codes to
eleven model grids, comprising a selection of
widely used sets of stellar evolution tracks and isochrones that have a range of
commonly adopted input physics. In applying several grid-pipeline
combinations, we capture implicitly in our final results the impact of
model dependencies from adopting different commonly-used grids, and differences in the
detail of the pipeline codes themselves.

 \subsubsection{Grid Pipelines}
 \label{sec:inputs}

Grid-based estimates of the stellar properties were returned by the
following pipeline codes:
 \begin{itemize}

 \item[--] The Yale-Birmingham (YB) \citep{Basu10, Basu12, Gai11};

 \item[--] The Bellaterra Stellar Properties Pipeline (BeSPP)
   (\citealt{Serenelli13} extended for asteroseismic analysis);

 \item[--] PARAM \citep{daSilva06, Miglio13};

 \item[--] RADIUS \citep{Stello09}; 

 \item[--] AMS \citep{Hekker13}; and

 \item[--] The Stellar Fundamental Parameters (SFP) pipeline
   \citep{Kallinger10, Basu11}.

 \end{itemize}

The YB pipeline was used with 5 different grids: models from the
Dartmouth group \citep{Dotter08} and the Padova group \citep{Marigo08, Girardi00}, 
the set of YY isochrones \citep{Demarque04}, 
a grid constructed using the Yale Stellar Evolution Code
(YREC; \citealt{Demarque08}) and described by \citet{Gai11} (we
refer to this set as YREC), and another set of models constructed with
a newer version of YREC with updated input physics
(we refer to this grid as YREC2) that has been described by Basu
et al. (2012). The Dotter et al. and Marigo et al. grids include
models of red-clump (RC) stars; YREC and YREC2 include only models of
He-core burning stars of higher mass, which do not go through the He
flash); while YY has no RC models. The BeSPP pipeline was run with two
grids.  The first grid is comprised of models constructed with the
GARSTEC code \citep{Weiss08} and the parameters of the grid
are described in \citet{SilvaAguirre12}. The second grid is
comprised of the BaSTI models of \citet{Pietrinferni04}, computed
for use in asteroseismic studies (see \citealt{SilvaAguirre13}). Both
grids include RC models. RADIUS was coupled to a grid constructed with
the ASTEC code (Christensen-Dalsgaard et al. 2008), as described in
\citet{Stello09} and \citet{Creevey12}, which does not include
RC models.

The codes above were all employed in the grid-based analysis of solar-type
\emph{Kepler} targets described in \citet{Chaplin14}, where
summary details of the physics employed in the grids may also be
found.

PARAM was run using a grid comprising models of the Padova group
\citep{Marigo08}, again including RC stars; further details may be
found in \citet{Miglio13}. AMS is based on an independent implementation of
the YB pipeline, and
was run using the BaSTI models of \citet{Pietrinferni04}. The SFP
pipeline was also coupled to BaSTI models. These grids include RC
models.

For this first analysis of the APOKASC red giants
an asteroseismic classification (i.e., RGB or RC) was not available
for many of the stars. Therefore, no \emph{a priori} categorisation
information was used in the grid-based searches. Although the cohort
evidently contains many RC stars, we nevertheless obtained some
results using grids comprised of only RGB models, to test the impact
of neglecting the red clump. However, as explained below, our final
results are produced using only those grids that included RC models.

 \subsubsection{Results from Grid-based Analyses}
 \label{sec:res}

\begin{figure*}\plotone{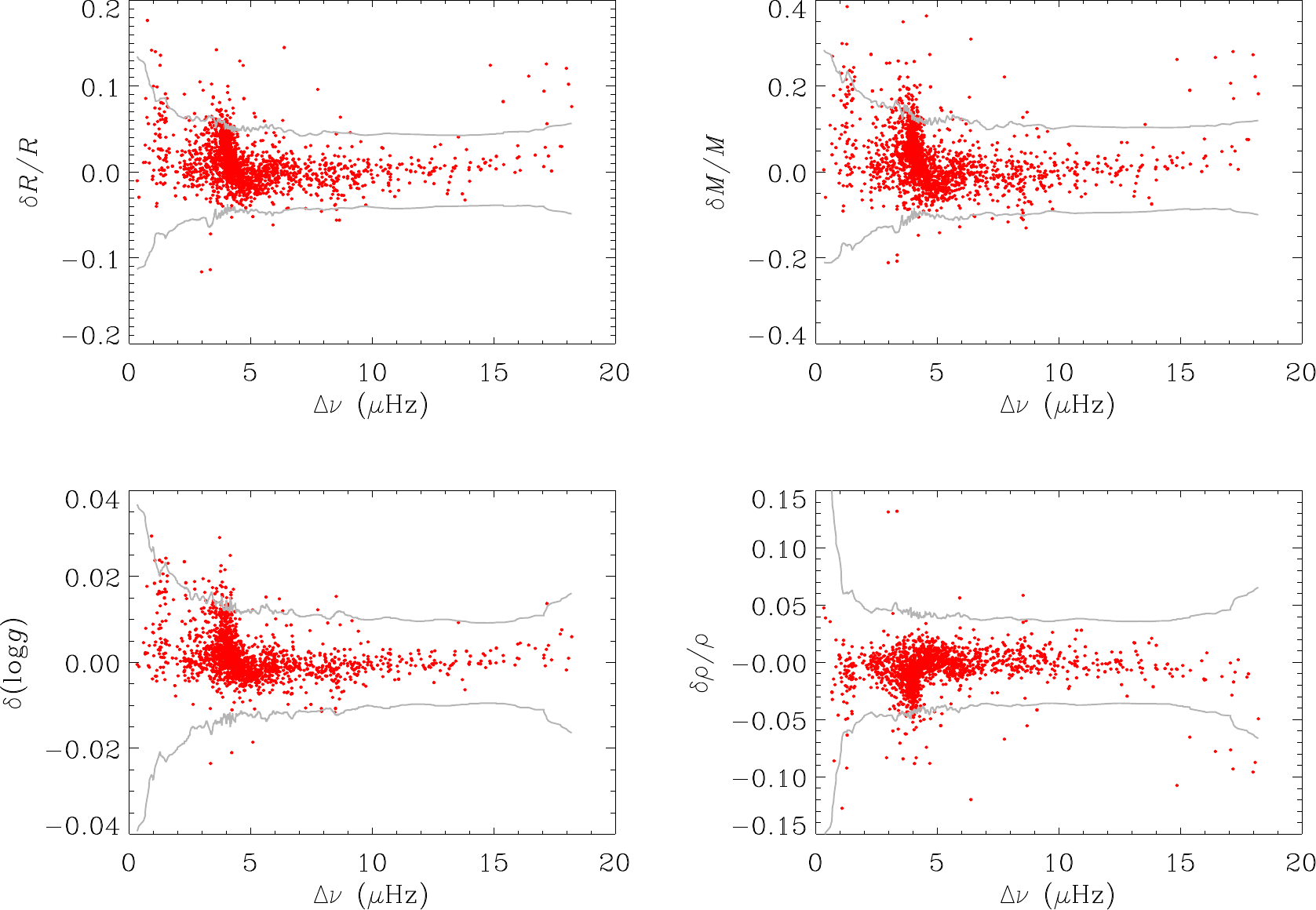}
  \caption{Fractional differences, as a function of
    $\Delta\nu$, in estimated properties returned by YB/Dotter and
    BeSPP/BaSTI. The results are shown for an analysis performed on the public
    data cohort, with the corrected ASPCAP $T_{\rm eff}$ and [Fe/H]
    values used as inputs. Gray lines mark the median $1\sigma$
    envelope of the grid-pipeline returned, formal uncertainties.
    These lines are included to help judge the \emph{typical}
    precision only.}
 \label{fig:diff1}
 \end{figure*}

Fig.~\ref{fig:diff1} is an example of the typical differences we see
in the estimated properties returned by different grid-pipeline
combinations, here those between YB/Dotter and BeSPP/BaSTI (the latter
chosen as the reference). Both sets comprise results from grids that
included RC models. Results are plotted from stars in the public data
cohort, with \dnu, $\nu_{\rm max}$, the revised ASPCAP $T_{\rm eff}$,
and [Fe/H] values used as inputs. We plot fractional
differences in $R$, $M$ and $\rho$ and absolute differences in
$\log g$. Gray lines mark envelopes corresponding to the median of
the $1\sigma$ uncertainties returned by all grid pipelines. Medians
were calculated in 10-target batches sorted on \dnu. These lines are
included to help judge the \emph{typical} precision only;
uncertainties in the results of individual targets may of course be
slightly different. Similar trends to those present here are seen
in results on the pixel and cluster target cohorts, and in results
from using the raw ASPCAP $T_{\rm eff}$ and [Fe/H] scales as
inputs.


On the whole, the differences lie  within the median formal uncertainty
envelopes, which is encouraging, i.e., the scatter between different
grid-pipeline combinations tends to be smaller than the typical
intrinsic, formal uncertainties returned by those pipelines. However,
we do see clear excess scatter centered on $\Delta\nu \simeq 4\,\rm
\mu Hz$, which corresponds to the location of RC stars. A significant
fraction of our target sample lies in this region. Further analysis
presented below suggests that this is genuine extra scatter, and not a
sampling effect (see Fig.~\ref{fig:fin2a_1} in Section~\ref{sec:cat}
and accompanying discussion). The presence of this scatter evidently
reflects the difficulty of discriminating between RC and RGB models
when no \textsl{a priori} categorisation is used as input, as was the
case here.

Differences with respect to the reference results of BeSPP/BaSTI tend,
not surprisingly, to be more pronounced for grids which did not
include RC models. Grid-pipeline combinations with no RC models appear
to compensate for the absence of the clump by the inclusion of
high-mass (lower-age) RGB solutions, which are not present in the RC
sets.  Although at the lowest masses the results for grids with and
without RC models are similar, the mapping to age is of course
different: grid-pipeline combinations with no RC models yield older
solutions at the \emph{same} mass. 


The lack of \textsl{a priori} information on the evolutionary state
has significant implications for our ability to return not only robust
estimates of the absolute ages, but even accurate measures of the
relative ages of the cohort, i.e., the relative chronology will be
scrambled if an RC star is incorrectly matched to a model of an RGB star, or
vice versa. Indeed, the problem is even more subtle. The
grid-based codes compute a likelihood for every model that is a
reasonable match (within several sigma) to the observables. Estimated
properties are returned from the distributions formed by these
likelihoods, i.e., the analysis is probabilistic in nature.  Without
information on the evolutionary state, the distribution functions may
be comprised of a mix of RC and RGB information. That is why, for now,
we do not provide ages explicitly in the catalog (although the reader
may compute their own ages from the masses and metallicities
provided). \citet{Bedding11} noted that the oscillation spectra of
the giants can be used to distinguish between RC and RGB stars, and we 
discussed methods earlier in the text that could provide automated
estimation of evolutionary state for many targets in the sample.  
Unfortunately, our initial screening procedure only produced
evolutionary state diagnostics for $\sim 25 \%$ of the targets.  We are
in the process of developing more efficient tools that will provide
measurements for an even larger fraction of the sample.
 Thus although we
do not have the information now, the desired evolutionary information
will be available to help construct the next version of the catalog.

\begin{figure*}\plotone{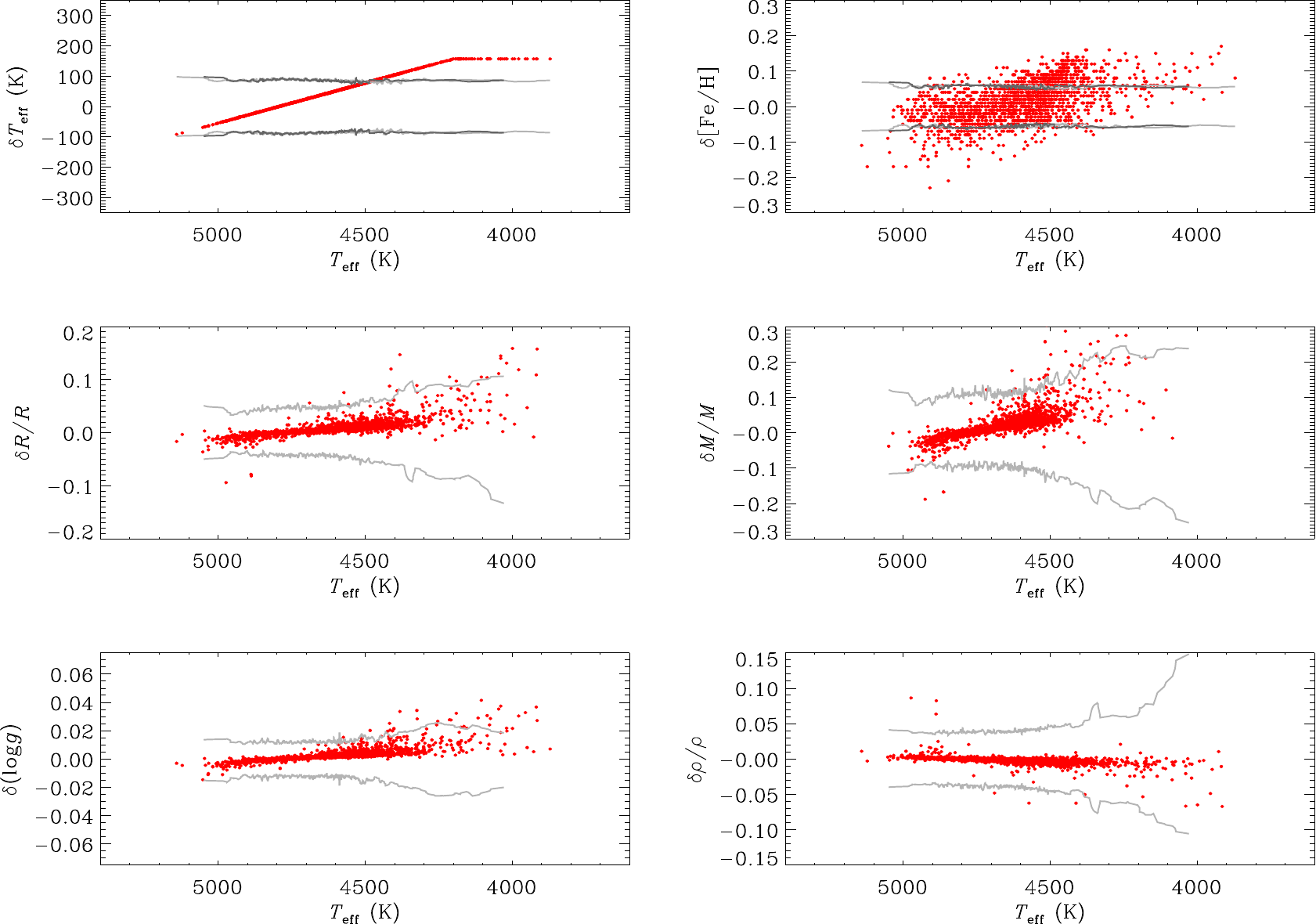}

\caption{Fractional differences in estimated properties returned by
  BeSPP/BaSTI, for analyses performed on the public data cohort with the
  two different $T_{\rm eff}$ and [Fe/H] inputs. Differences are
  defined as results with corrected ASPCAP scale inputs minus
  results with raw ASPCAP scale inputs. Gray lines mark the median
  $1\sigma$ envelope (over all pipelines) of the returned, formal
  uncertainties. The top panels show the absolute $T_{\rm eff}$ and
  [Fe/H] differences between the two sets of input parameters, while
the bottom four panels record the impact on the derived asteroseismic properties.
 The impact on the separate mass and radius measurements (in the middle panels)
 is larger than the impact on the surface gravity and mean density (in the bottom panels).}

\label{fig:diffteffin}
\end{figure*}

Fig.~\ref{fig:diffteffin} shows the impact on the public data cohort
results of switching from one set of ASPCAP $T_{\rm eff}$ and [Fe/H]
inputs to the other. Results for the pixel and cluster cohorts show
similar trends. The top left-hand panel plots the corrected ASPCAP
$T_{\rm eff}$ minus the raw ASPCAP $T_{\rm eff}$, showing the
piece-wise correction that was applied to the raw temperatures to
yield the corrected scale. The lines follow the median $1\sigma$
envelopes of the uncertainties.  The top
right-hand panel presents the corrected ASPCAP [Fe/H] minus the raw
ASPCAP [Fe/H]. The other panels display the fractional differences in
estimated properties returned by BeSPP/BaSTI, in the sense corrected
ASPCAP minus raw ASPCAP. As in the previous figures, gray lines
mark the median $1\sigma$ envelopes (over all pipelines) of the
returned, formal uncertainties.


With reference to the asteroseismic scaling relations, the trends
revealed in Fig.~\ref{fig:diffteffin} may be understood largely in
terms of the changes to the temperature scale. The relations imply
that, all other things being equal, $M \propto T_{\rm eff}^{1.5}$, $R
\propto T_{\rm eff}^{0.5}$ and $g \propto T_{\rm eff}^{0.5}$, while
the seismic estimates of $\rho$ are not affected by the change to
$T_{\rm eff}$. The plotted property differences are thus seen to
reflect, to good approximation, the trend in $T_{\rm eff}$, although
this clearly does not explain all the differences, i.e., there is also
the impact of the changes in [Fe/H] to consider (which are for example
apparent in the small differences seen in the estimates of $\rho$).

 \subsubsection{Asteroseismic Catalog Properties and Uncertainties}
 \label{sec:cat}

We provide tables of estimated properties for each of the raw
ASPCAP and corrected ASPCAP scales (see Tables 4 and 5 below).  For both sets of inputs, the
properties in the catalog are those that were returned by
BeSPP/BaSTI. Its results lay closest to the median over all
grid-pipelines and targets. By choosing one grid-pipeline to provide
the final properties we avoid mixing results that are subject to
different input physics and pipeline methodology. We instead opted to
reflect those differences in the quoted final uncertainties, by taking
into account the scatter between results returned by the different
grid-pipeline combinations. Our approach is 
therefore similar to that adopted by \citet{Chaplin14} for asteroseismic
dwarfs and subgiants.  We emphasize that in this consolidation we used
only results from grids that included RC models.  The 
BeSPP/BaSTI models did not include convective overshoot or semi-convection
and incorporated mass loss with an efficiency $\eta=0.4$.  These choices did
not have a major impact on the derived masses and radii.

\begin{figure*}\plotone{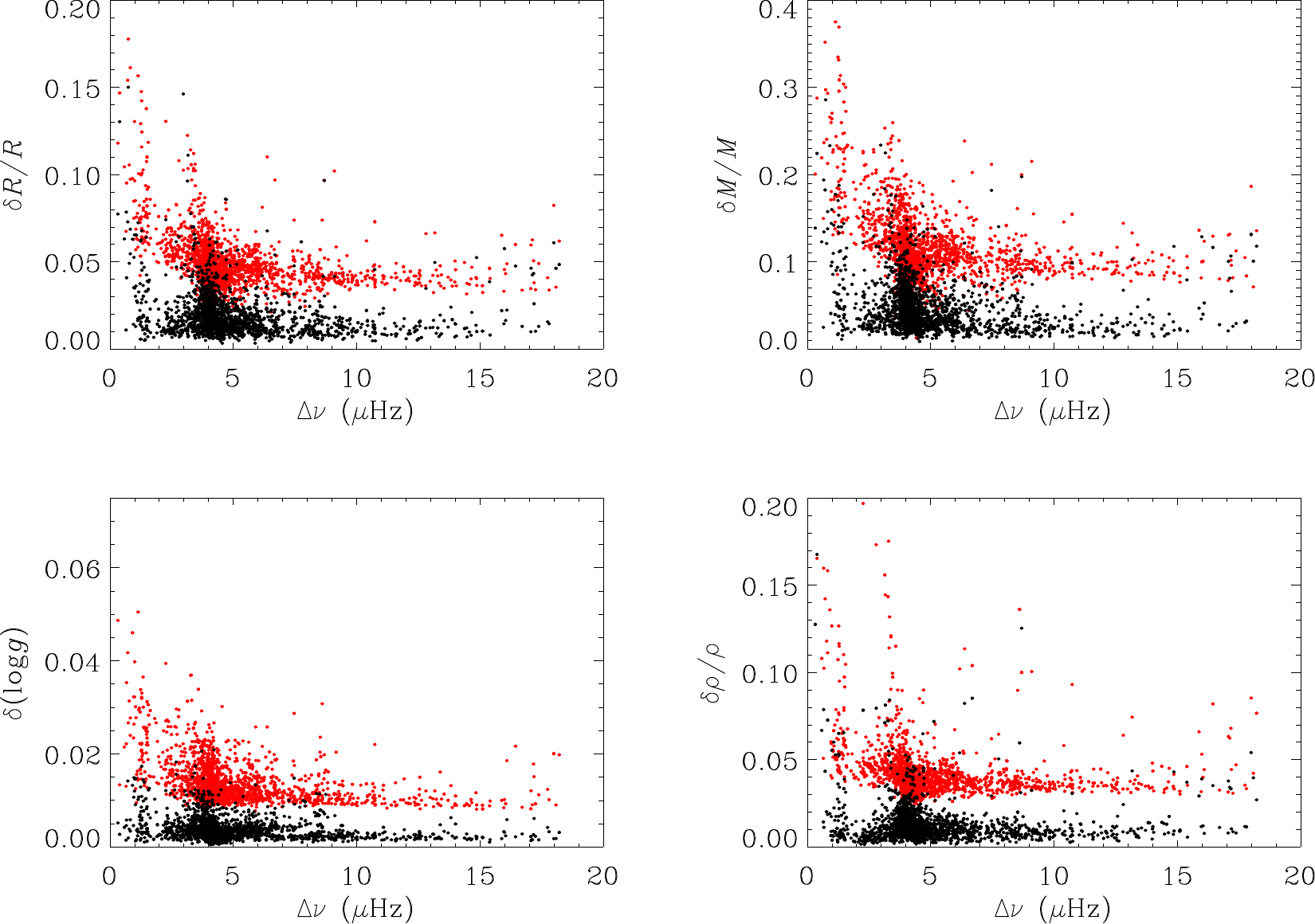}

 \caption{Median formal uncertainties over all grid pipelines (for
   grids with RC models; red symbols), for the public data cohort
   (corrected ASPCAP scale inputs); and the standard deviations
   (scatter) between them (black symbols), both as a function of
   \dnu.  The excess scatter on the far left reflects the difficulty of inferring
precise astroseismic properties for luminous giants, while the excess scatter close to
$4 \rm \mu Hz$ is caused by ambiguity between RC and RGB solutions.}

\label{fig:fin2a_1}
\end{figure*}

Fig.~\ref{fig:fin2a_1} plots the median formal uncertainties returned
by grid-pipeline combinations including RC models only (red symbols), again for results
from the public data cohort using the corrected ASPCAP scale inputs. For
each property of every star we calculated the standard deviation of
the results returned by the various grid-pipeline combinations. These
estimates of the scatter are plotted in black.  The median standard
deviations are approximately 4\,\% in mass, 1\,\% in radius,
0.003\,dex in $\log\,g$, and 1\,\% in density.

The plots demonstrate clearly that the scatter between the grid-pipelines is
typically smaller than the formal uncertainties returned by any one
pipeline. Moreover, both the scatter \textsl{and} the formal
uncertainties are larger at the $\Delta \nu$ where we find RC stars. This result
again emphasizes the extra challenges posed for the grid searches in
this part of the parameter space in the absence of an \emph{a priori}
categorisation.


\begin{figure*}\plotone{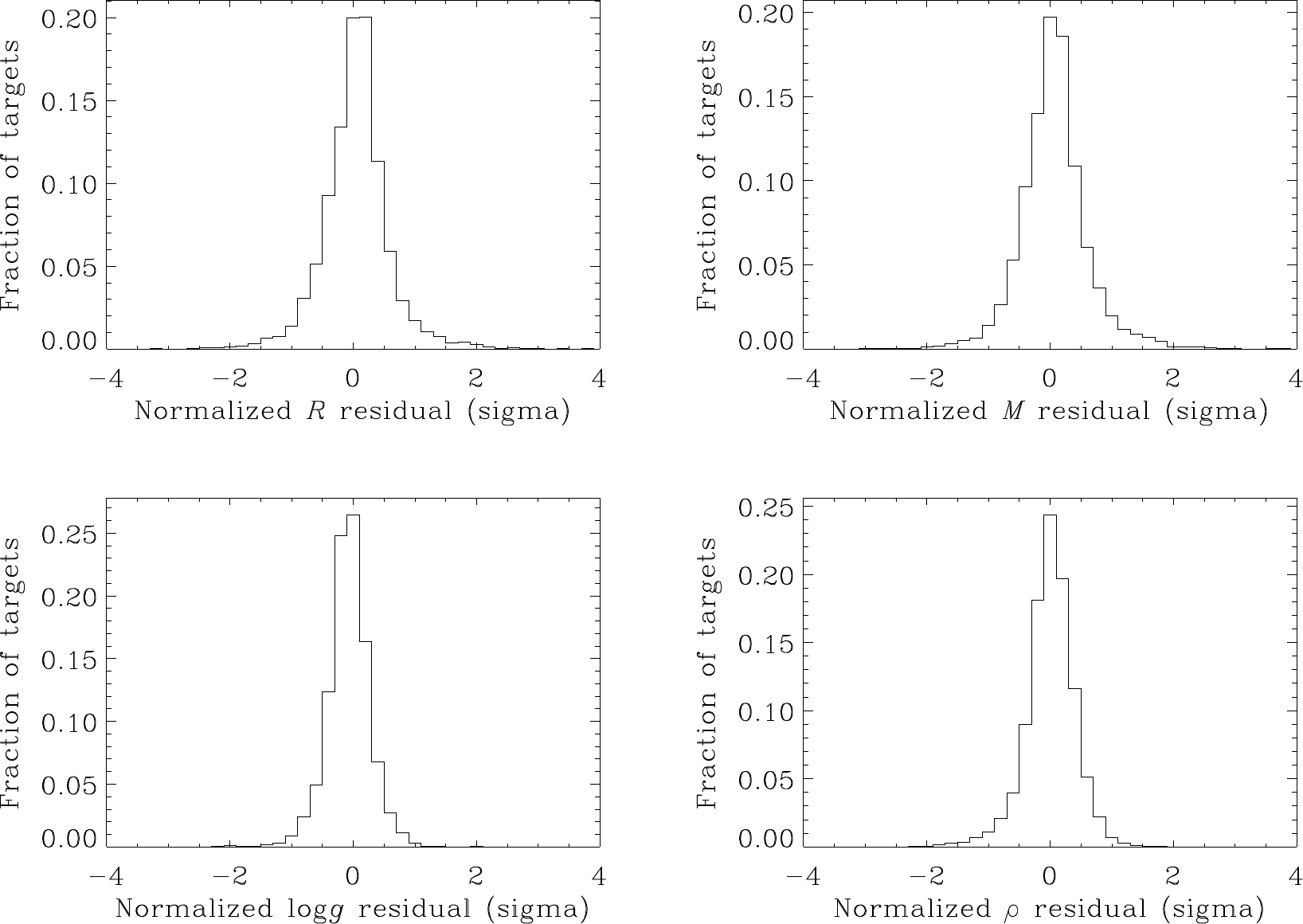}

\caption{Histograms, for each property, of uncertainty-normalized
  residuals over all pipeline-grid combinations (omitting BeSPP/BaSTI)
  and all stars in the pubic data cohort (corrected ASPCAP scale
  inputs). The plotted residuals were calculated with respect to the
  BeSPP/BaSTI results.}

\label{fig:superdis}
\end{figure*}

The above measures of scatter were combined in quadrature with the
individual formal uncertainties returned by BeSPP/BaSTI to yield the
final uncertainties on the estimated properties. Distributions of the
scatter between pipelines are shown in Fig.~\ref{fig:superdis}. The
plotted histograms (here, for the same cohort of results as the previous
figures) were constructed by computing residuals for each pipeline-grid with
respect to BeSPP/BaSTI, and normalizing each residual by the median
property uncertainty given by the pipelines for that star. We then
accumulated residuals for all stars in the cohort, and binned the
residuals to give the plotted ``super distributions''.  The most
striking aspect of all the histograms is their Gaussian-like
appearance (see similar results for solar-type stars in \citealt{Chaplin14}). 
Our median final uncertainties are approximately 12\,\% in
mass, 5\,\% in radius, 0.01\,dex in $\log\,g$, and 3\,\% in
density. These uncertainties will typically be higher for those stars
which have $\Delta\nu \simeq 4\,\rm \mu Hz$ (i.e., RC stars).


Finally, some targets appeared in both the pixel
and public-data samples. In such cases we adopted the properties
estimated from the higher-quality, longer pixel lightcurves to be the
catalog properties. Estimated properties for the cluster stars all
came from analyses of the cluster data.

A question that might be legitimately be asked is why should we resort to
grid-based modeling, which makes our results model dependent,
when the direct use of Equation~\ref{eq:dnu} and 
Equation~\ref{eq:numax} could, in principle, produce model-independent results. 
The issue is that the scaling relations 
assume that all values of $T_{\rm eff}$ are possible for a star of a 
given mass and radius and are unconstrained by the equations of
stellar structure and evolution. Thus it is quite possible that observational
errors in $T_{\rm eff}$, $\Delta\nu$ and $\nu_{\rm max}$ can produce 
estimated values of mass and radius inconsistent with the theory of stellar structure and evolution.
The grid-based method takes this constraint into account implicitly, as well as
including metallicity information not used in the scaling relations.


\begin{figure*}\plottwo{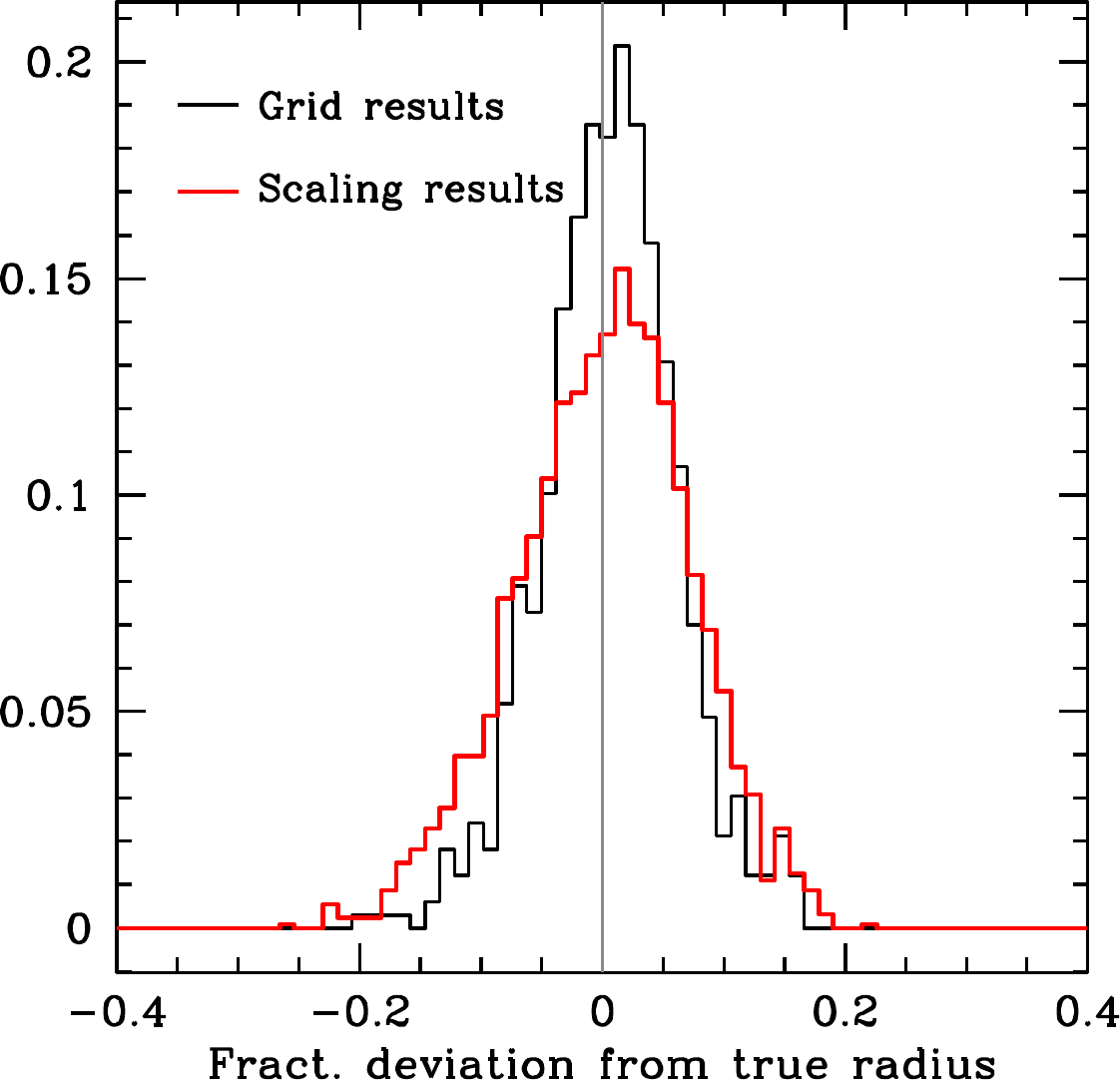}{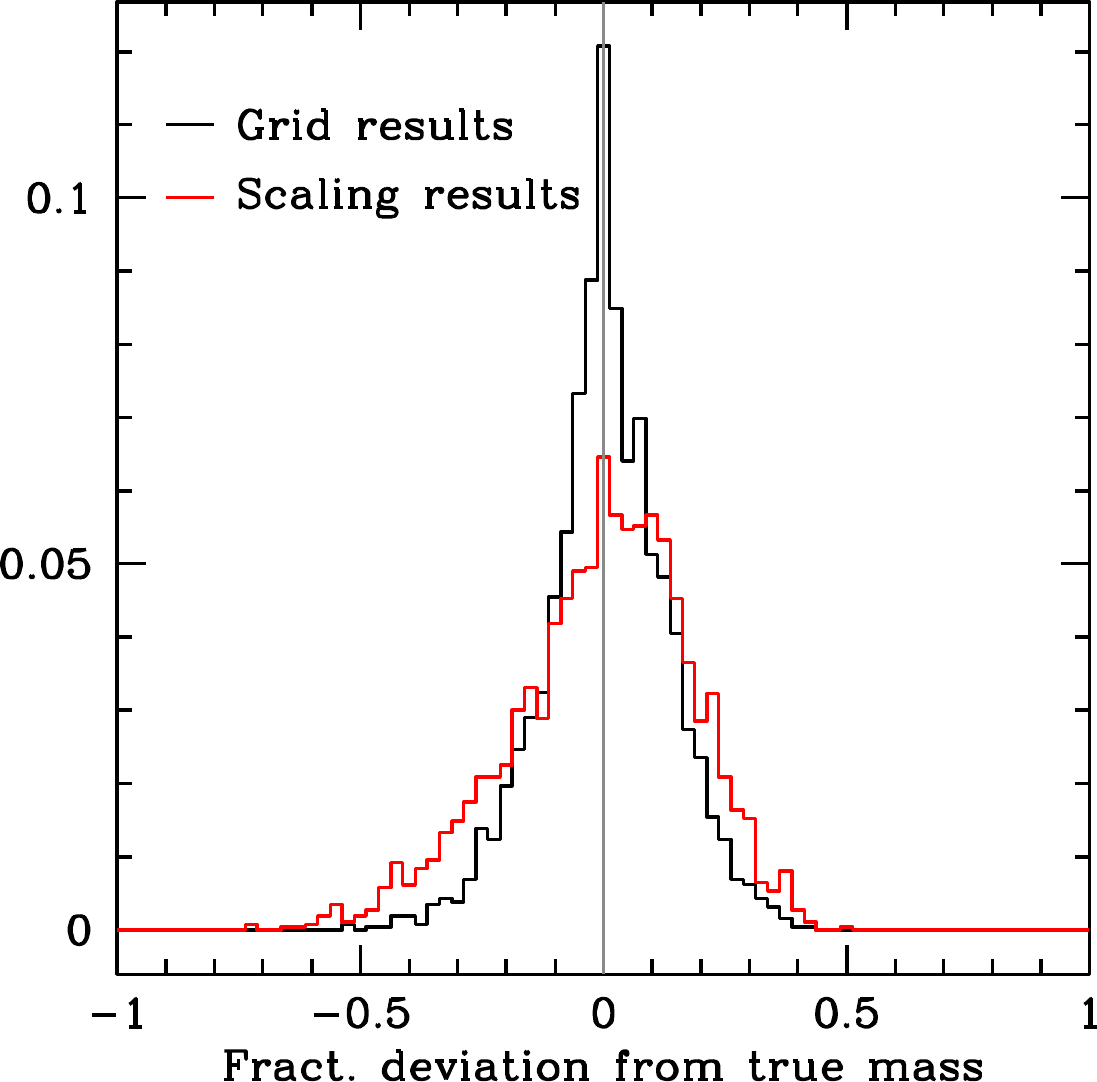}

\caption{Deviations between true and measured radii(left) and mass(right) for simulated data
using scaling relations alone (red) and using grid modeling (black).
The simulated dataset is taken from Gai et al. 2011.}

\label{fig:simradius}
\end{figure*}


Properties of asteroseismic grid-modeling results have been studied
in detail by Gai et al. (2011).  They showed that propagated errors are
in general smaller in grid-based results than those inferred from propagating scaling relations. 
The Gai et al. work 
dealt with simulated stars over a large part of the HR diagram, and the 
gain obtained by grid-based modeling may not persist
in the narrow temperature range occupied by red giants. To test this hypothesis, we have
taken the Gai et al. simulated sample and selected all objects with
$\Delta\nu < 20 \mu$Hz as likely red giants.  In Figure
~\ref{fig:simradius}
we show histograms of the fractional difference between the true and inferred
radius and mass for these stars obtained with both the scaling relations 
and through a grid-based search. In both cases, the distribution
of the deviation between the true and the inferred properties is sharper for the grid-based
method. Although differences between the distributions are not as 
large as in the case of main-sequence or subgiant stars, the difference, particularly in 
mass, is large enough that it is worth performing grid modeling for our sample to reduce our random uncertainties.  As discussed above, grid modeling is more challenging
in the $\log g$ regime where both the RGB and RC are present, and RC models and stars are not included in this comparison.

\section{The APOKASC Catalog and Its Properties}

\subsection{The APOKASC Catalog}

We present our data in three tables.  Table 3 contains basic information on the stars, and our asteroseismic parameters for the raw and 
corrected temperature scale are given in Tables 4 and 5.  We begin in Table 3 with the information in 
common for both sets of spectroscopic inputs.  We begin with star identifiers and positional information: the KIC (first column) and 2MASS (second column) IDs of the targets; RA and Dec (J2000), Galactic longitude l and latitude b (all in degrees)  are presented in columns 3-6.  We present two different $T_{\rm eff}$  measurements (in K) estimates using the GHB09 methodology in columns 7 and 8: one with the KIC extinction and the other with zero extinction.  
The uncertainties reflect
random errors in the $J-K$ color used for the temperature estimates.  This is distinct from the published effective temperatures presented in \citet{Pinsonneault12}, which
used either SDSS filters or the \citet{Casagrande10} IRFM temperatures from 2MASS colors.  For each star that passed our internal consistency checks, we then provide
the mean asteroseismic properties $\Delta\nu$ and $\nu_\mathrm{max}$ (in $\mu$Hz) and their uncertainties in columns 9 and 10.  The final column includes the targeting flags used to select the stars for APOGEE observations.  Stars included in GO proposal $\#40033$ are labelled GO; stars targeted as public giants are PUBLIC (note that this includes almost all targets not newly observed for the GO program.) A small number of asteroseismic ''dwarfs'' with long cadence frequency measurements are labeled as SEIS DWARF.

Our detection efficiency was a function of surface gravity, which reflects physical and sampling effects that make obtaining reliable
results more challenging for some targets than for others.  The highest gravity stars had oscillation frequencies close to the Nyquist frequency, which can create model-dependent results.
Low gravity stars have fewer modes and their interpretation is also more model dependent.  The stars for which we did not obtain consistent results were clustered around these two categories.
For the analysis of the full DR12 catalog we intend to reanalyze and report results for all targets with detections, but we did not proceed with the full analysis of such objects in our
initial run.

\begin{deluxetable}{lllllllllll}
\tabletypesize{\tiny}
\setlength{\tabcolsep}{0.04in} 
\tablecaption{APOKASC Catalog Basic Data}
\tablehead{\colhead{KIC ID} & \colhead{2MASS ID} & \colhead{RA} & \colhead{Dec.} & \colhead{Gal. l} & \colhead{Gal. b} & \colhead{GHB T$_\mathrm{eff}$} & \colhead{GHB T$_\mathrm{eff}$} & \colhead{$\Delta\nu$} & \colhead{$\nu_\mathrm{max}$} & \colhead{Star}\\ \colhead{} & \colhead{} & \colhead{} & \colhead{} & \colhead{} & \colhead{} & \colhead{E(B-V)=KIC} & \colhead{E(B-V)=0} & \colhead{($\mu$Hz)} & \colhead{($\mu$Hz)} & \colhead{Flags}}
\startdata
10907196 & J18583782+4822494 & 284.658 & 48.380  &   78.402  &   18.804    &   4969 $\pm$   79  &  4808  $\pm$ 74  &  4.67  $\pm$  0.13 &  44.12  $\pm$ 0.96  &  KASC,PUBLIC \\
10962775 & J18582020+4824064 & 284.584 & 48.402  &   78.405  &   18.857    &   5009 $\pm$   81  &  4819  $\pm$ 75  &  4.11  $\pm$  0.09 &  35.12  $\pm$ 0.86  &       PUBLIC \\
11177749 & J18571019+4848067 & 284.292 & 48.802  &   78.735  &   19.174    &   4769 $\pm$  105  &  4612  $\pm$ 99  &  4.11  $\pm$  0.10 &  34.18  $\pm$ 0.79  &         KASC \\
11231549 & J18584464+4857075 & 284.686 & 48.952  &   78.977  &   18.981    &   4728 $\pm$   80  &  4574  $\pm$ 75  &  3.40  $\pm$  0.08 &  30.25  $\pm$ 0.69  &         KASC \\
11284798 & J18582108+4901359 & 284.588 & 49.027  &   79.028  &   19.067    &   4349 $\pm$   70  &  4201  $\pm$ 65  &  1.46  $\pm$  0.04 &   9.50  $\pm$ 0.25  &  KASC,LUMINOUS \\
\enddata\label{table:Results1}
\end{deluxetable}

In Table 4 we present the asteroseismic properties derived from the uncorrected spectroscopic parameters; we refer to this set of measurement as Scale 1.  Table 5 has an identical format \
except that it was derived using the corrected spectroscopic
parameters, and we refer to these measurements as Scale 2.  We start with the KIC ID, the input $T_{\rm eff}$ in K, and [M/H] (the logarthmic iron to hydrogen ratio relative to the Sun) for the relevant spectroscopic scale.  
We then present the mass (in solar units), radius (in solar units), surface
gravity (log base 10 in cgs units), and mean density relative to that of the Sun.  See Section 4 for the methodology used to derive the uncertainties.  Although one could in principle construct mean densities and 
surface gravities from the masses and radii, we solved for these independently.  The tabulated density and gravity values are close, but therefore not identical, to those which could be inferred from the separate $M$ and $R$.

\begin{deluxetable}{lrrrrrr}
\tablecaption{Scale 1 Asteroseismic Results}
\tablehead{ \colhead{KIC ID} & \colhead{ASPCAP T$_\mathrm{eff}$} & \colhead{ASPCAP [M/H]} & \colhead{Mass} & \colhead{Radius} & \colhead{$\log g$} & \colhead{$\rho$} \\  \colhead{} & \colhead{(Raw)(K)} & \colhead{(Raw)} & \colhead{M$_\odot$} & \colhead{R$_\odot$} & \colhead{} & \colhead{$solar units$} }
\startdata
10907196 & $4721 \pm 87 $ &  $-0.08\pm 0.06$ & $1.47^{+0.16}_{-0.19}$ & $10.77^{+0.42}_{-0.64}$ & $2.543^{+0.011}_{-0.011}$ & $0.00119^{+0.00005}_{-0.00005}$ \\
10962775 & $4716 \pm 94 $ &  $-0.25\pm 0.06$ & $1.21^{+0.13}_{-0.12}$ & $10.93^{+0.47}_{-0.48}$ & $2.444^{+0.011}_{-0.012}$ & $0.00093^{+0.00003}_{-0.00003}$ \\
11177749 & $4591 \pm 83 $ &  $ 0.03\pm 0.05$ & $1.08^{+0.12}_{-0.11}$ & $10.51^{+0.49}_{-0.45}$ & $2.426^{+0.011}_{-0.011}$ & $0.00093^{+0.00003}_{-0.00003}$ \\
11231549 & $4451 \pm 86 $ &  $-0.06\pm 0.06$ & $1.51^{+0.17}_{-0.15}$ & $13.31^{+0.58}_{-0.56}$ & $2.368^{+0.011}_{-0.011}$ & $0.00064^{+0.00002}_{-0.00002}$ \\
11284798 & $4126 \pm 85 $ &  $-0.04\pm 0.06$ & $1.22^{+0.17}_{-0.14}$ & $21.76^{+1.14}_{-1.05}$ & $1.848^{+0.013}_{-0.013}$ & $0.00012^{+0.00001}_{-0.00001}$ \\
\enddata\label{table:Results2}
\end{deluxetable}

\begin{deluxetable}{lrrrrrr}
\tablecaption{Scale 2 Asteroseismic Results}
\tablehead{ \colhead{KIC ID} & \colhead{ASPCAP T$_\mathrm{eff}$} & \colhead{ASPCAP [M/H]} & \colhead{Mass} & \colhead{Radius} & \colhead{$\log g$} & \colhead{$\rho$} \\   \colhead{} & \colhead{(Corrected)} & \colhead{(Corrected)} & \colhead{} & \colhead{} & \colhead{} & \colhead{} }
\startdata
  10907196   &  4740 $\pm$   87 &  -0.08  $\pm$  0.06 &  $1.50^{+0.13}_{-0.20}$ & $10.88^{+0.33}_{-0.72}$ & $2.543^{+0.011}_{-0.011}$ & $0.00119^{+0.00005}_{-0.00004}$ \\
  10962775   &  4736 $\pm$   94 &  -0.29  $\pm$  0.06 &  $1.21^{+0.13}_{-0.12}$ & $10.94^{+0.48}_{-0.45}$ & $2.444^{+0.011}_{-0.012}$ & $0.00093^{+0.00003}_{-0.00003}$ \\
  11177749   &  4644 $\pm$   83 &   0.06  $\pm$  0.05 &  $1.08^{+0.13}_{-0.11}$ & $10.54^{+0.50}_{-0.46}$ & $2.427^{+0.011}_{-0.010}$ & $0.00093^{+0.00003}_{-0.00003}$ \\
  11231549   &  4541 $\pm$   86 &  -0.03  $\pm$  0.06 &  $1.55^{+0.17}_{-0.16}$ & $13.46^{+0.61}_{-0.58}$ & $2.371^{+0.011}_{-0.011}$ & $0.00064^{+0.00002}_{-0.00002}$ \\
  11284798   &  4283 $\pm$   85 &   0.04  $\pm$  0.06 &  $1.29^{+0.16}_{-0.15}$ & $22.26^{+1.12}_{-1.07}$ & $1.854^{+0.012}_{-0.012}$ & $0.00012^{+0.00001}_{-0.00001}$ \\\enddata\label{table:Results3}
\end{deluxetable}

\subsection{APOKASC and the KIC Compared}

\begin{figure*}
 \plotone{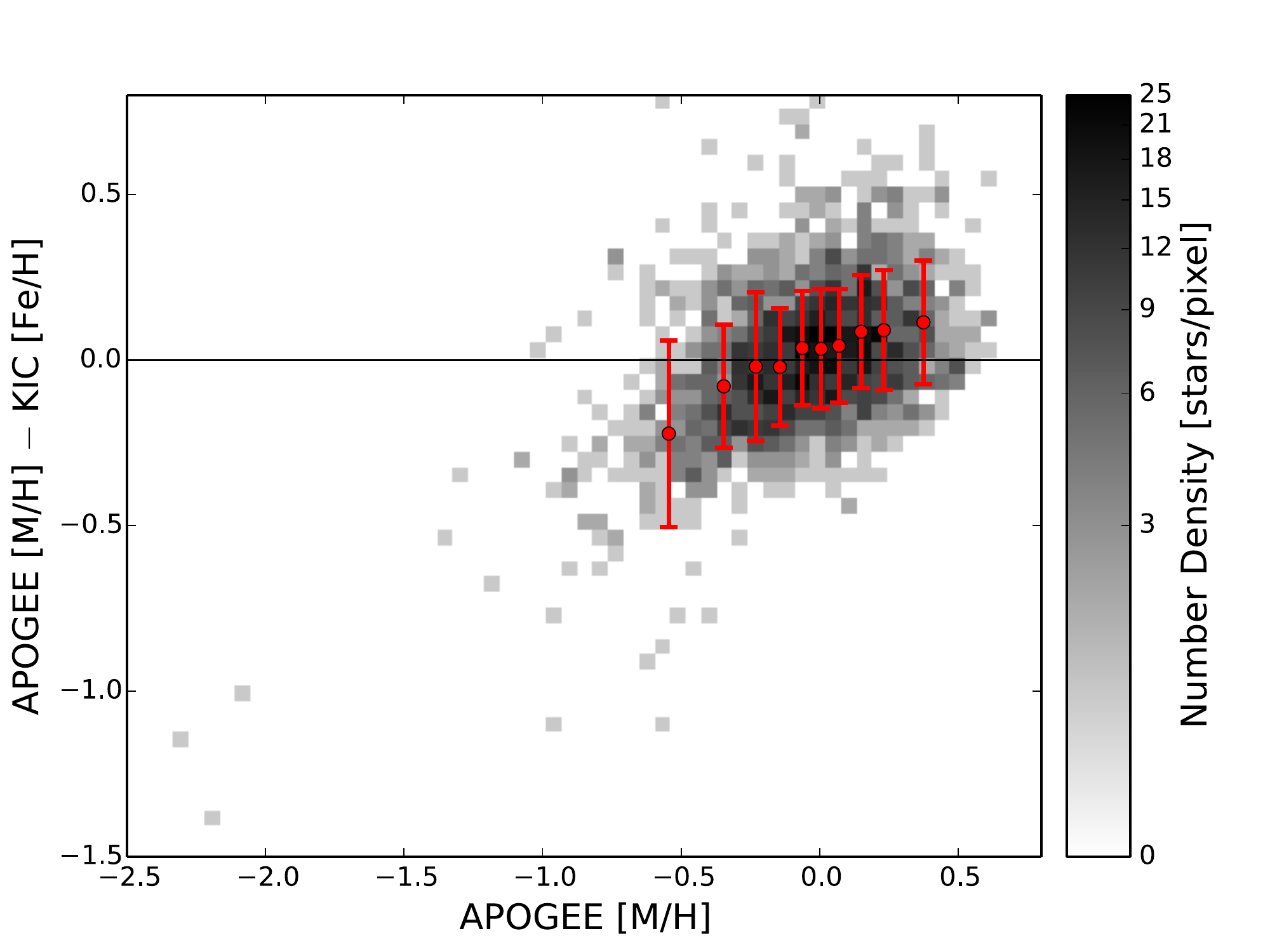}
  \caption{Logarithmic difference in metallicity between the APOKASC
and KIC metallicities as a function of APOKASC metallicity.  The points with error bars
are the means and standard deviations of the data in 10 ranked cohorts of APOKASC
metallicity.  The data were divided into 60 bins in metallicity and metallicity difference, covering
the range $-2.5$ to $+0.8$ and $-1.5$ to $+0.8$ dex, respectively,
and the logarithmic color coding (specified on the right) indicates the number of targets with those properties.
The correspondence between the KIC and spectroscopic results for giants is closer than that reported previously for
dwarfs.}
 \label{fig:FeHKIC}
 \end{figure*}

\begin{figure*}
 \plotone{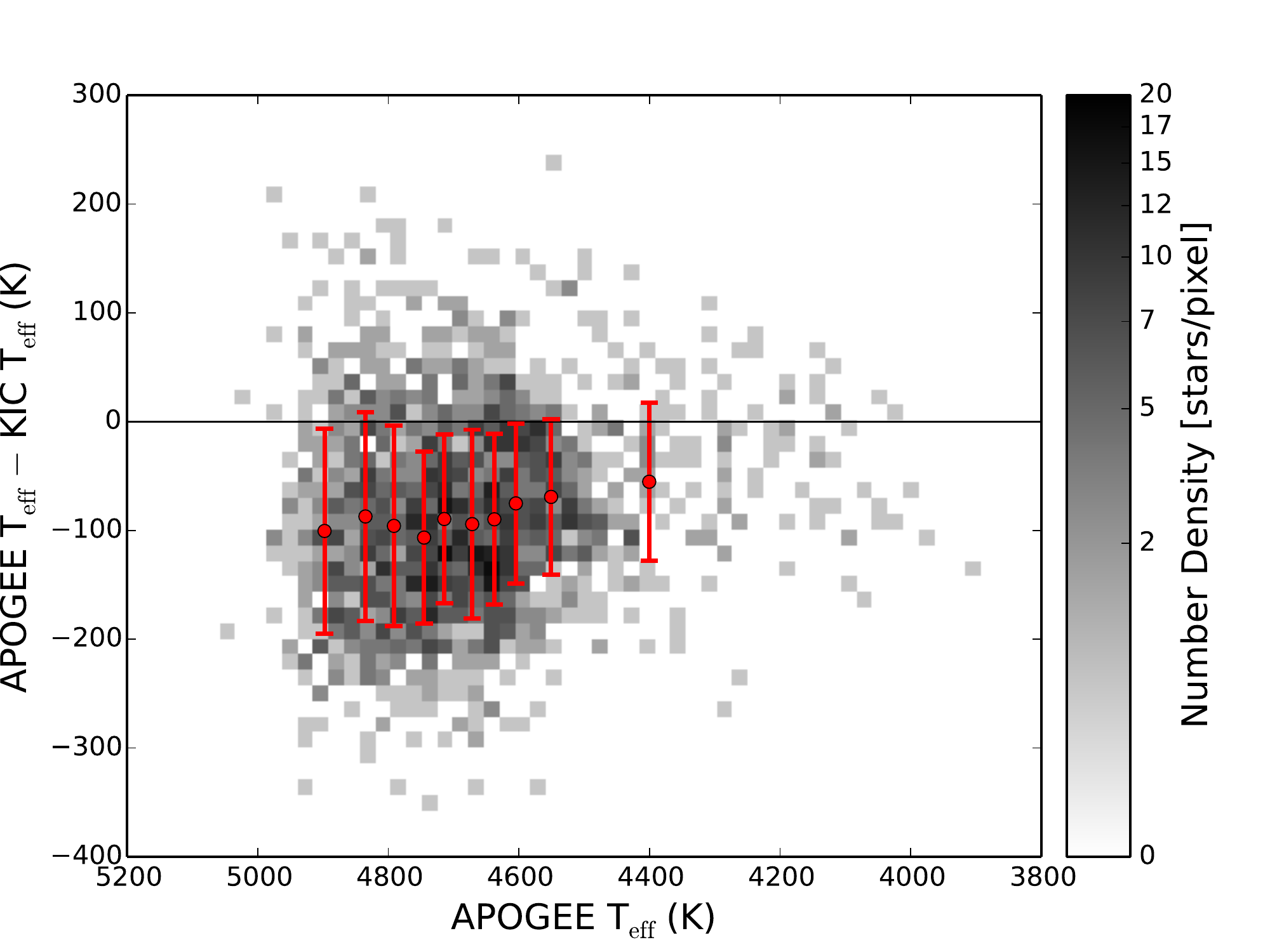}
  \caption{Difference in effective temperature between the APOKASC
and KIC values as a function of APOKASC $T_{\rm eff}$ (K).  The points with error bars
are the medians and median absolute deviations of the data in 10 ranked cohorts of APOKASC
$T_{\rm eff}$.  The data were divided into 60 bins in $T_{\rm eff}$ and $\Delta T_{\rm eff}$, covering the range 3800 K  to 5200 K
and $-400$ to $+300$ K, respectively, and the logarithmic color coding (specified on the right) indicates the number of 
targets with those properties.  The major difference between the two systems is a zero-point offset.}
 \label{fig:TeffKIC}
 \end{figure*}

\begin{figure*}
 \plotone{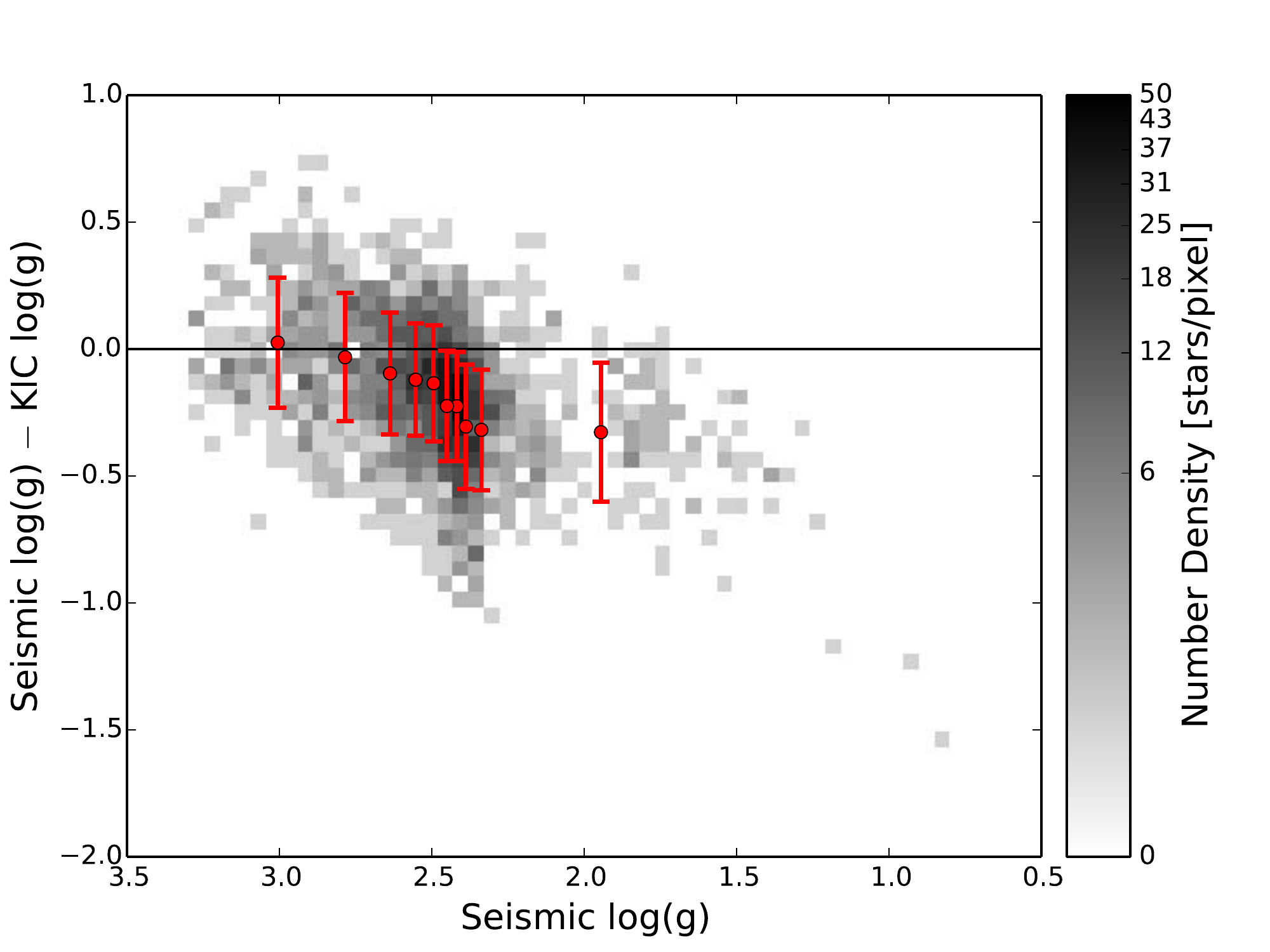}
  \caption{Logarithmic difference in surface gravity $\log g$ between the APOKASC
and KIC values as a function of APOKASC $\log g$.  The points with error bars
are the medians and median absolute deviations of the data in 10 ranked cohorts of APOKASC
log g.  The data was divided into 60 bins in $\log g$  and delta $\log g$, covering the range 0.5 to 3.5 and $-2.0$ to $+1.0$, respectively,
and the logarithmic color coding (specified on the right) indicates the number of targets with those properties.}
 \label{fig:LoggKIC}
 \end{figure*}

A natural point of reference for our work are the stellar parameters in the Kepler Input
Catalog (KIC), which were based on photometry.  Fig.~\ref{fig:FeHKIC} displays the binned
metallicity differences between our spectroscopic metallicities and those in the
KIC inferred from photometry as a function of APOKASC metallicity.  The average
metallicities are surprisingly close, with a mean difference of 0.006 dex and 
a standard deviation of 0.22 dex.  However, there are significant metallicity trends in the
difference between the two, in the sense that the KIC metallicity scale is compressed relative to the
APOGEE one. The effective temperature scale is also systematically cooler than that derived from the
KIC, demonstrated in Fig.~\ref{fig:TeffKIC}. The offset ($-86$ K on average) and dispersion
 ($84$ K) are modest.  The temperature scale differences are sensitive
to the adopted extinction model and to the color-temperature calibration, an issue that we
 return to below.  Finally, the differences in surface gravity are illustrated in Fig.~\ref{fig:LoggKIC}.
The KIC surface gravities have a large scatter of $0.265$ dex relative to our asteroseismic values, with a mean
difference of $-0.175$ dex.


\begin{figure*}
 \plotone{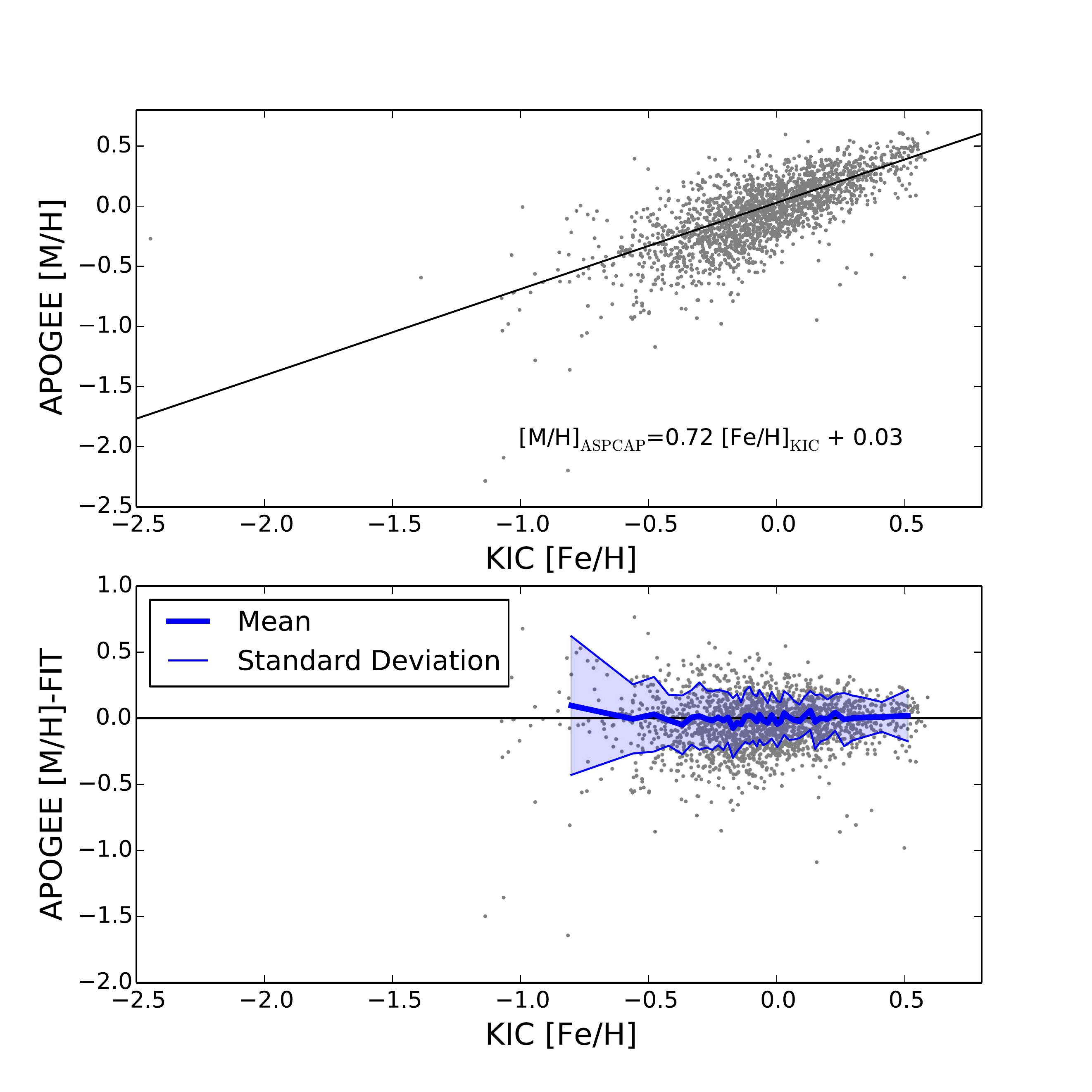}
  \caption{Results of a linear fit between the KIC and ASPCAP corrected metallicities. A strong metallicity trend is seen in
the residuals, indicating a lower precision for photometric metallicity estimates in metal-poor stars.}
 \label{fig:FeHKICfit}
 \end{figure*}

\begin{figure*}
 \plotone{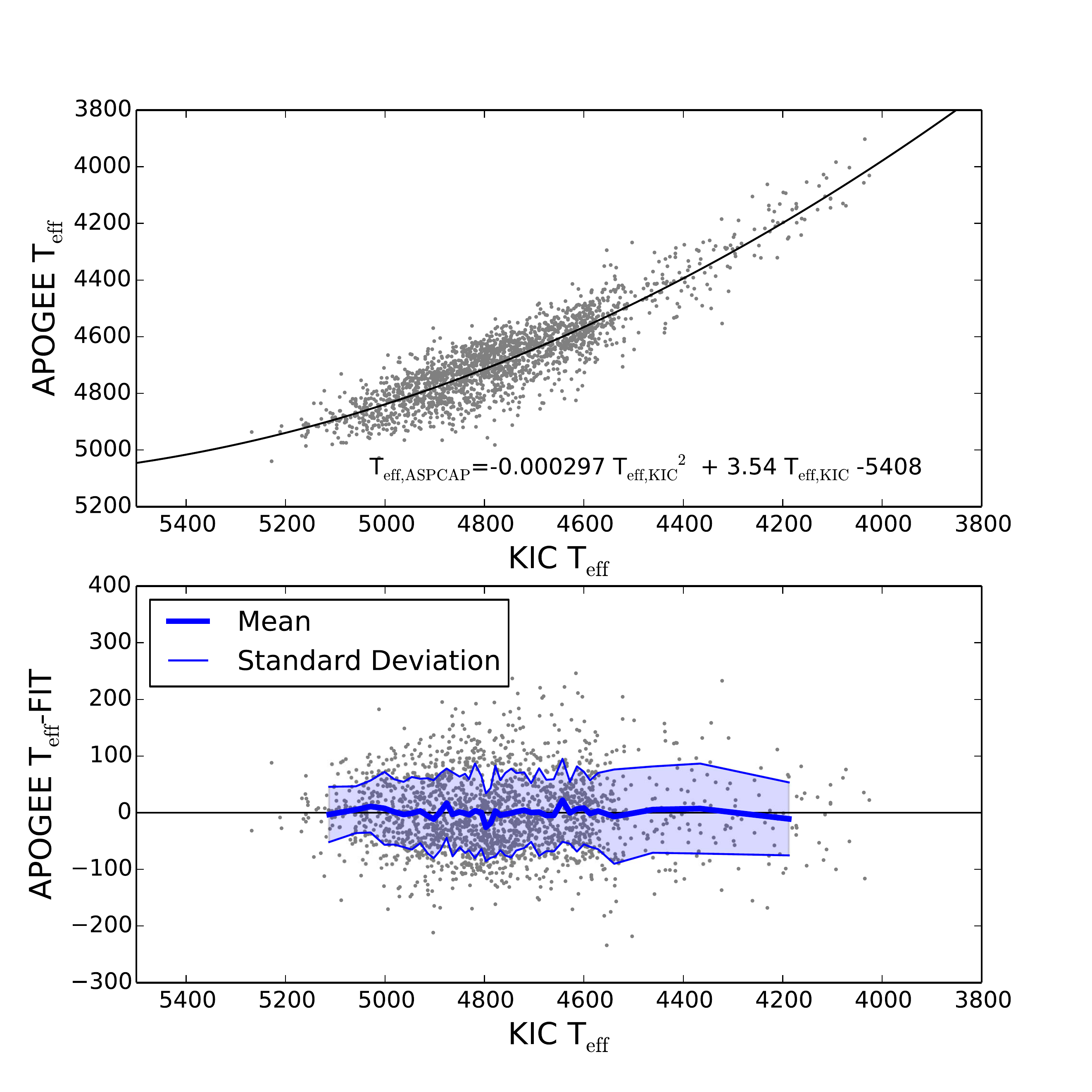}
  \caption{Results of a quadratic fit between the KIC and ASPCAP corrected $T_{\rm eff}$ values.  There are only weak trends in uncertainties with
effective temperature.  Hot stars, which are mostly in the secondary clump in our sample, are largely
responsible for the curvature in the fitting function.}
 \label{fig:TeffKICfit}
 \end{figure*}

\begin{figure*}
 \plotone{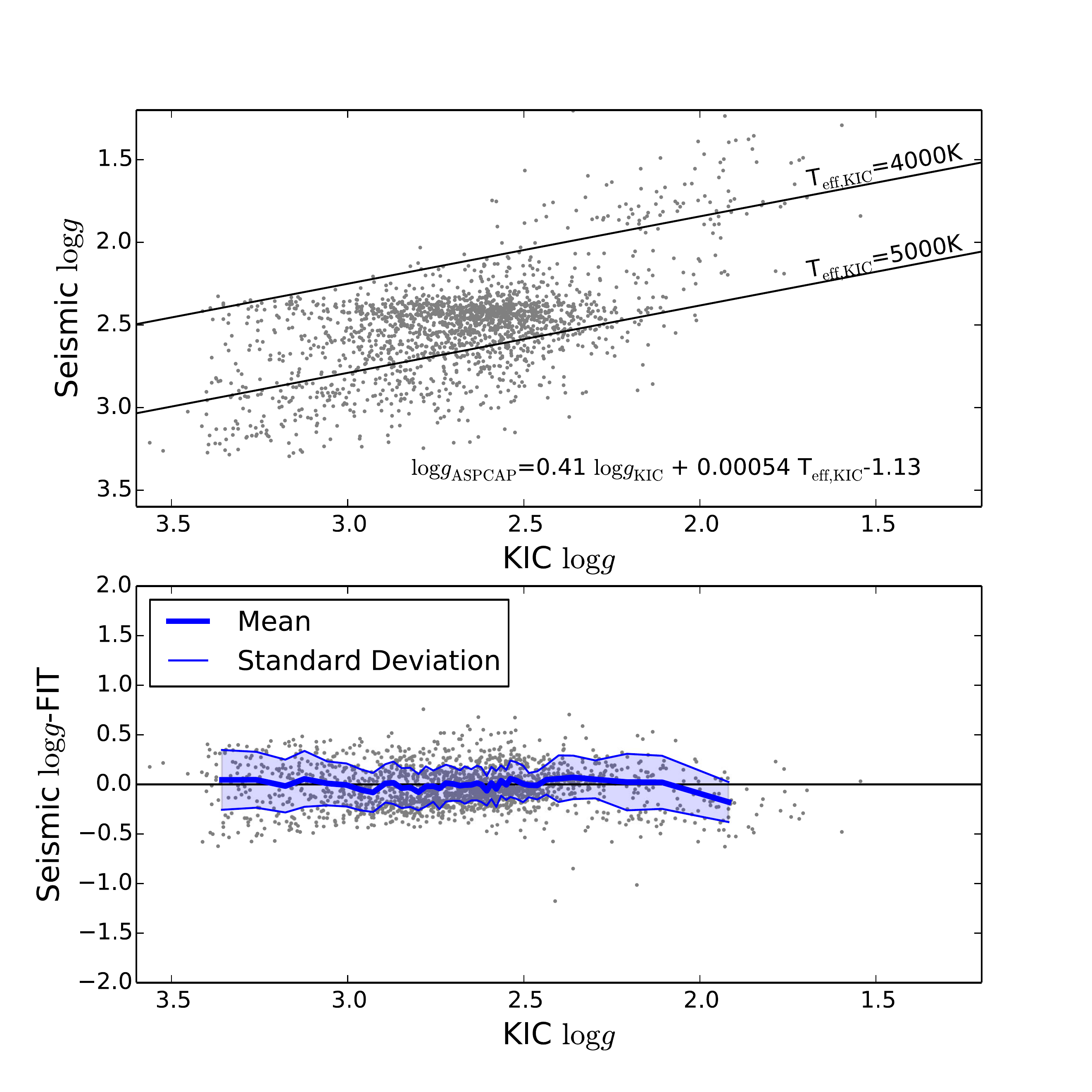}
  \caption{Results of a bilinear fit between the KIC and ASPCAP corrected $T_{\rm eff}$ values.  Our best fit relations for cool and
hot giants correspond to the two lines on the figure.}
 \label{fig:LoggKICfit}
 \end{figure*}

Motivated by these results we derived fitting functions mapping the KIC parameters onto the spectroscopic system.
We emphasize that these results are calibrated for red giants only, and different relations may well apply for
dwarfs or subgiants.  The fits between the KIC parameters and the corrected ASPCAP [M/H], $T_{\rm eff}$ and asteroseismic $\log g$ 
are shown in Figures ~\ref{fig:FeHKICfit}, ~\ref{fig:TeffKICfit}, and ~\ref{fig:LoggKICfit}, respectively.  Our fitting formulae
correcting the KIC values to the spectroscopic scale are

  \begin{equation} 
  {[M/H]_{ASPCAP}} = 
  {0.72*[Fe/H]_{KIC}+0.03}.
  \label{eq:FeHKIC}
  \end{equation}

  \begin{equation} 
  {T_{\rm eff,ASPCAP}} = 
  {0.000297*T_{\rm eff,KIC}^2 + 3.54*T_{\rm eff,KIC}-5408}.
  \label{eq:TeffKIC}
  \end{equation}

  \begin{equation} 
  {\log g_{ASPCAP}} = 
  {0.41*\log g_{KIC} + 0.00054*T_{\rm eff,KIC}-1.13}.
  \label{eq:LoggKIC}
  \end{equation}

The top panel of each figure shows the best-fit line compared to the data, with the function in the lower right. 
The dark and light blues curves are the weighted mean and magnitude of the standard deviation respectively (computed by dividing the data into 40 bins containing equal numbers of stars). 
A linear function was a good fit for Equation~\ref{eq:FeHKIC}.  Curvature in the differences at high temperatures made a quadratic function a better fit than a linear one for Equation~\ref{eq:TeffKIC}, 
and correlations between the gravity offsets, $T_{\rm eff}$, and $\log g$ drove our choice of functional form for Equation~\ref{eq:LoggKIC}.   
The scatter around the KIC [Fe/H]-corrected ASPCAP [M/H] relationship increases toward lower metallicity stars; this result is not surprising for a photometric system,
which tends to lose sensitivity for more metal-poor objects.  \citet{Dong13} performed a related exercise comparing LAMOST and KIC metallicities for dwarfs.  In their case, they reported a comparable
dispersion in the relative metallicities but a different functional form for the fit.  This discrepancy is likely to be tied to the difference between the photometric calibrations for giants and dwarfs. Similar
effects are also seen for $T_{\rm eff}$  and \logg, in the sense that the dwarf and giant offsets between the KIC and comparison samples are not the same.  


\begin{figure*}
 \plotone{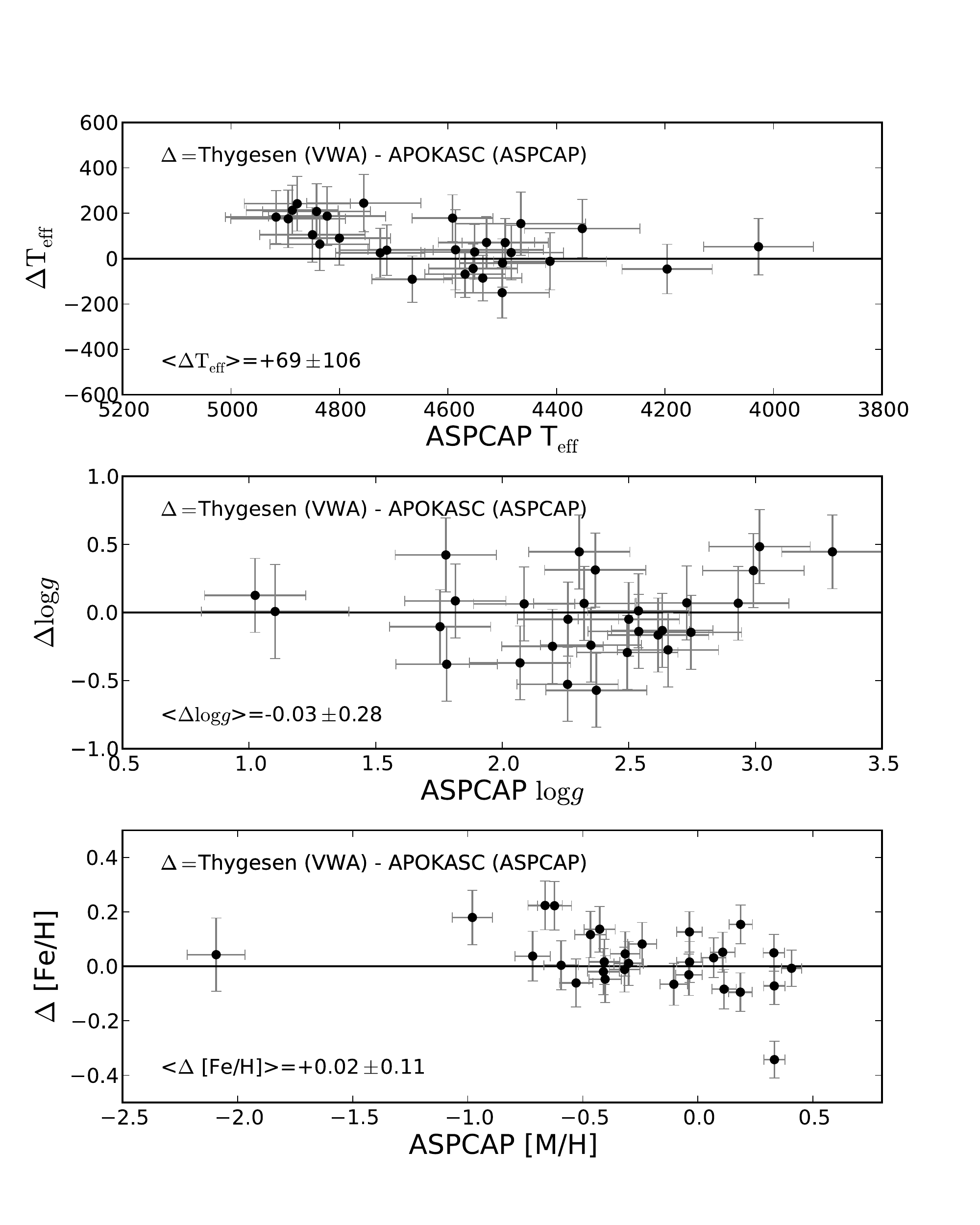}
  \caption{Temperature, surface gravity, and [Fe/H] differences, respectively, for stars in common between APOGEE 
and Thygesen et al. APOGEE internal errors are used.  Thygesen individual error bars were not directly reported, so they were
inferred to be half (in quadrature) of the differences between their optical results and external literature comparisons. 
The derived dispersions between optical and IR spectra are indicated on the figures and are consistent with this uncertainty measurement.}
 \label{fig:optvsir}
 \end{figure*}

Our final external check is against an sample of results from optical spectroscopy obtained by \citet{Thygesen12}. Those authors compared their spectroscopic results 
against both asteroseismic surface gravities and the KIC.  That paper reported
a close correspondence between the mean KIC and spectroscopic properties, with spectroscopic minus KIC differences $\Delta$ $T_{\rm eff}$ , $\Delta$ $\log g$, and $\Delta [Fe/H]$ of 3 K ($\sigma = 105$ K), 
$-0.003$ dex ($\sigma = 0.67$ dex) and $0.003$ dex ($\sigma = 0.50$ dex) respectively.  Relative to the asteroseismic surface gravities, \citet{Thygesen12} reported a
smaller offset than the one that we obtain ($-0.05$ dex) with a slightly larger scatter ($0.30$ dex).   
Further insight can be obtained from the stars in common between the two samples (Figure ~\ref{fig:optvsir}).  The ASPCAP and optical spectroscopy results are in good agreement in both metallicity and surface gravity, and disagree
only in effective temperature (at the $86$ K level).  The close agreement between our results and those of \citet{Thygesen12} for stars in common between the two samples
indicates that there is no global $\log g$ offset between the two spectroscopic methods.  Because our sample is large and the uncertainties in the asteroseismic gravities are small, our mean offset of -0.175 dex between the KIC
and asteroseismic scales is highly
statistically significant, which differs from the -0.05 dex offset reported by \citet{Thygesen12}.  Since the two methods agree in their results
for the stars in common, the origin of this difference is likely to be in sampling effects (e.g. our sample includes stars where the asteroseismic and KIC surface gravities simply disagree more on average than a typical
star in the smaller Thygesen sample).   The zero point offset between the corrected spectroscopic temperatures and 
those of Thygesen is comparable in magnitude and sign with that inferred relative to the photometric GHB09 scale. The temperature zero-point offsets, however, appear to have minimal impact on the metallicities or surface gravities.


\subsection{Global Properties of the APOKASC Sample}

\begin{figure*}
 \plotone{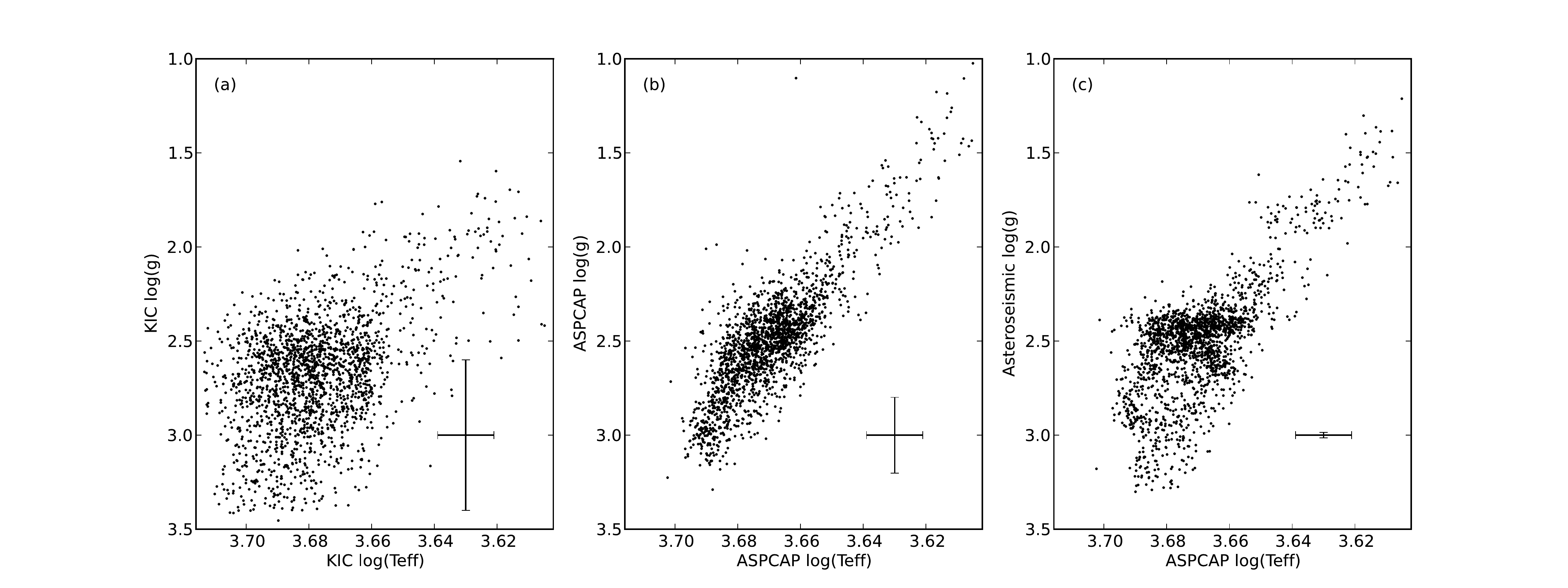}
  \caption{Our catalog stars in $\log g$ - $T_{\rm eff}$ space for three different methods.  The original, purely photometric system for the KIC is in 
the left panel.  A purely spectroscopic system, corresponding to the parameters released in
DR10, is in the center panel.  Our published parameters, which differ by the addition of asteroseismic surface gravities, produces the right panel.  
The stellar populations in the field snap into focus as we provide additional
information.}
 \label{fig:focus}
 \end{figure*}

The combination of asteroseismic and spectroscopic data adds new dimensions to traditional stellar population studies.
As an example, we present three different HR diagrams for the stars in our sample in Figure ~\ref{fig:focus}.
The left panel reflects the photometric parameters inferred from the KIC.  With the addition of spectroscopic data distinct
features begin to emerge, in particular a prominent red clump.  However, with the addition of asteroseismic surface gravities,
fine structure can clearly be seen in a field population without parallax data.  Both the red clump and the secondary red clump
can be clearly distinguished, and the red giant branch bump (typically detected only in star clusters) is clearly visible.


\begin{figure*}
 \plottwo{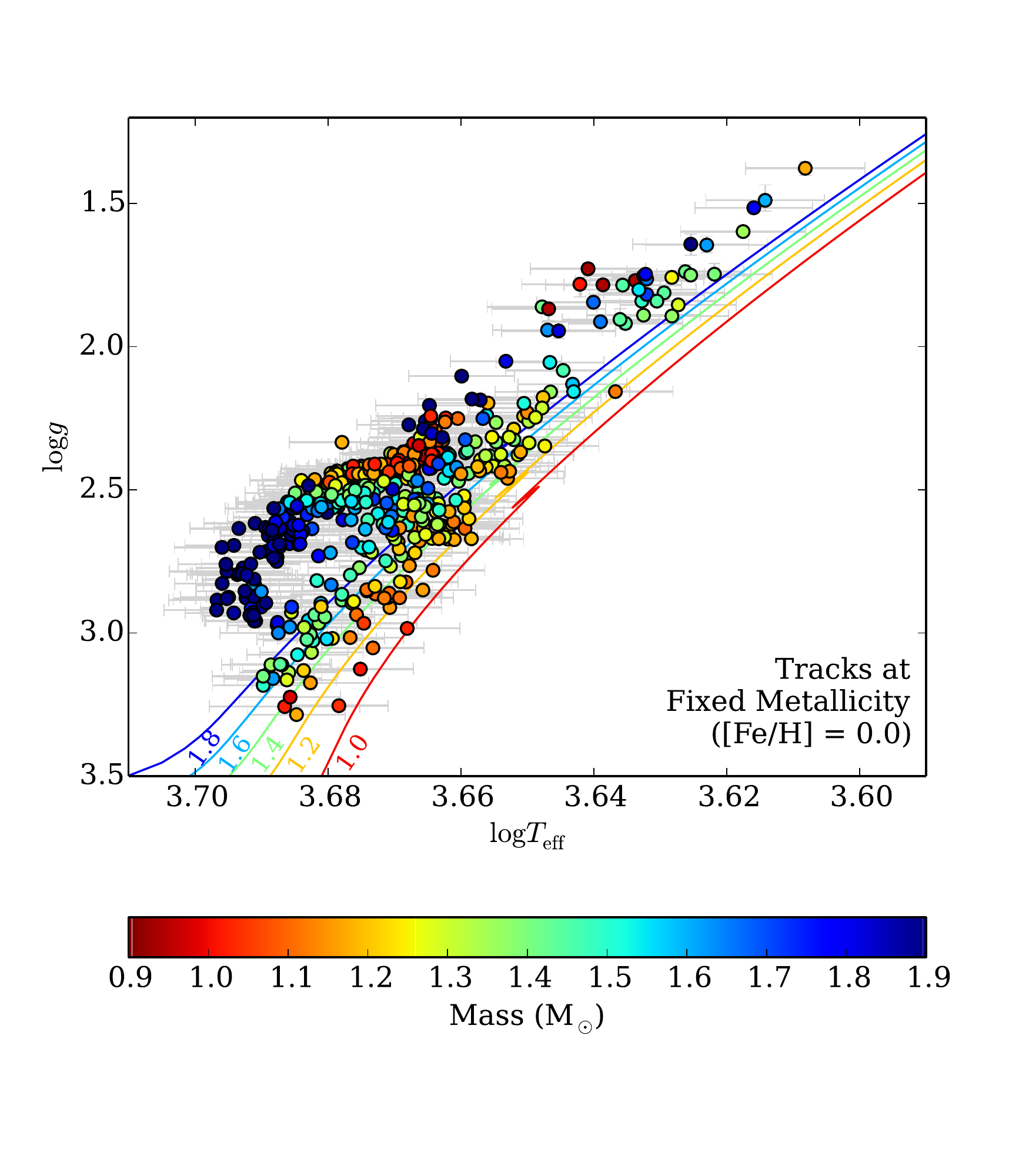}{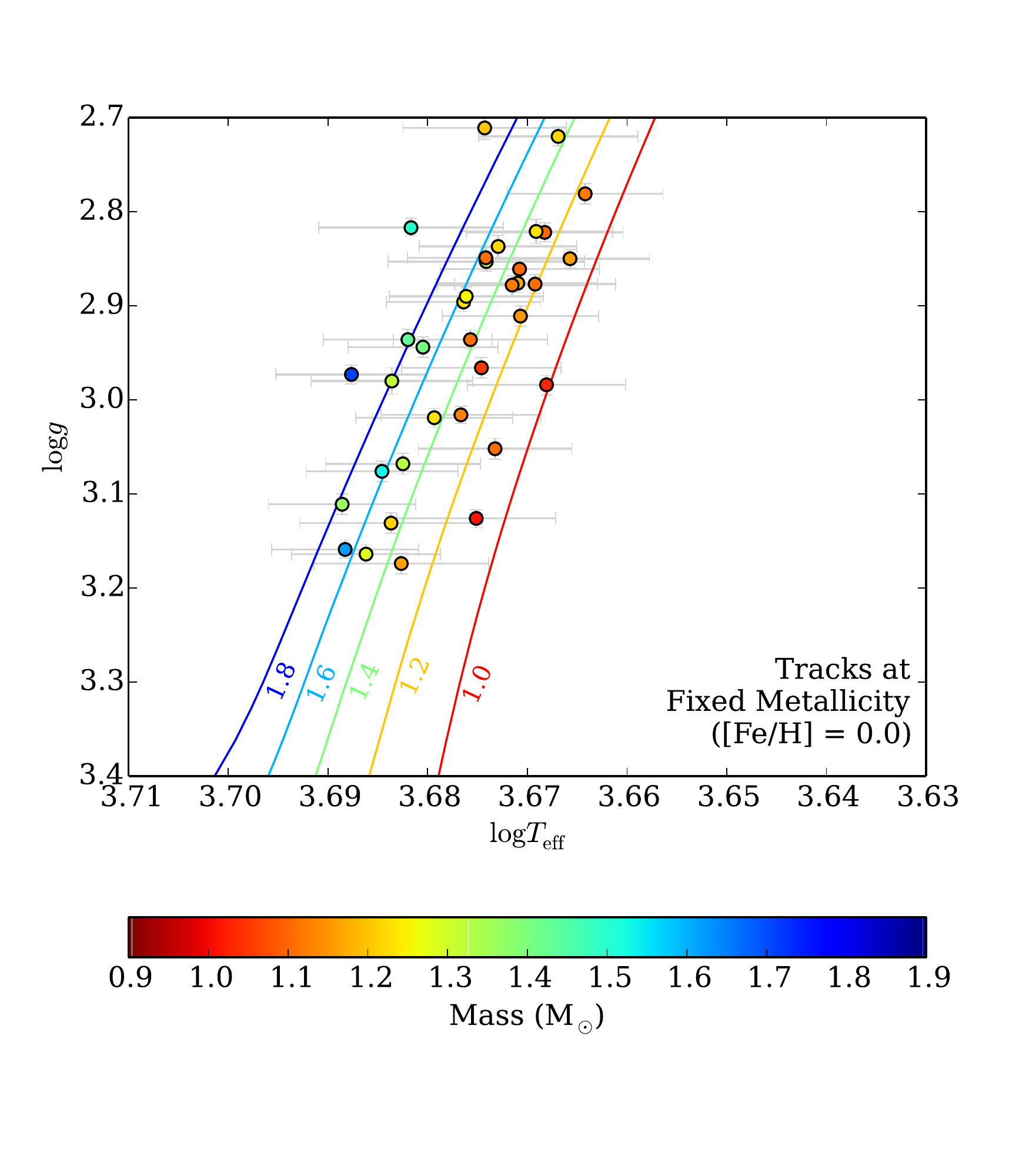}
  \caption{Left: HR diagram for all APOKASC stars in the metallicity bin $-0.1<$ [Fe/H] $ < +0.1$, 
color coded by seismic mass such that lighter colors correspond to less massive stars. 
For comparison, YREC tracks with [Fe/H]=0 and Mass=1.0, 1.2, 1.4, 1.6 M$_\odot$ are shown. Right: 
Same as for the left panel, except that the sample is restricted to seismically classified red giant 
branch stars on the lower red giant branch. There  is some evidence for a systematic offset in the magnitude of the 
mass trend for the lowest mass targets, but this could reflect the finite age of the disk; see text.}
 \label{fig:FeHTrends}
 \end{figure*}

\begin{figure*}
 \plottwo{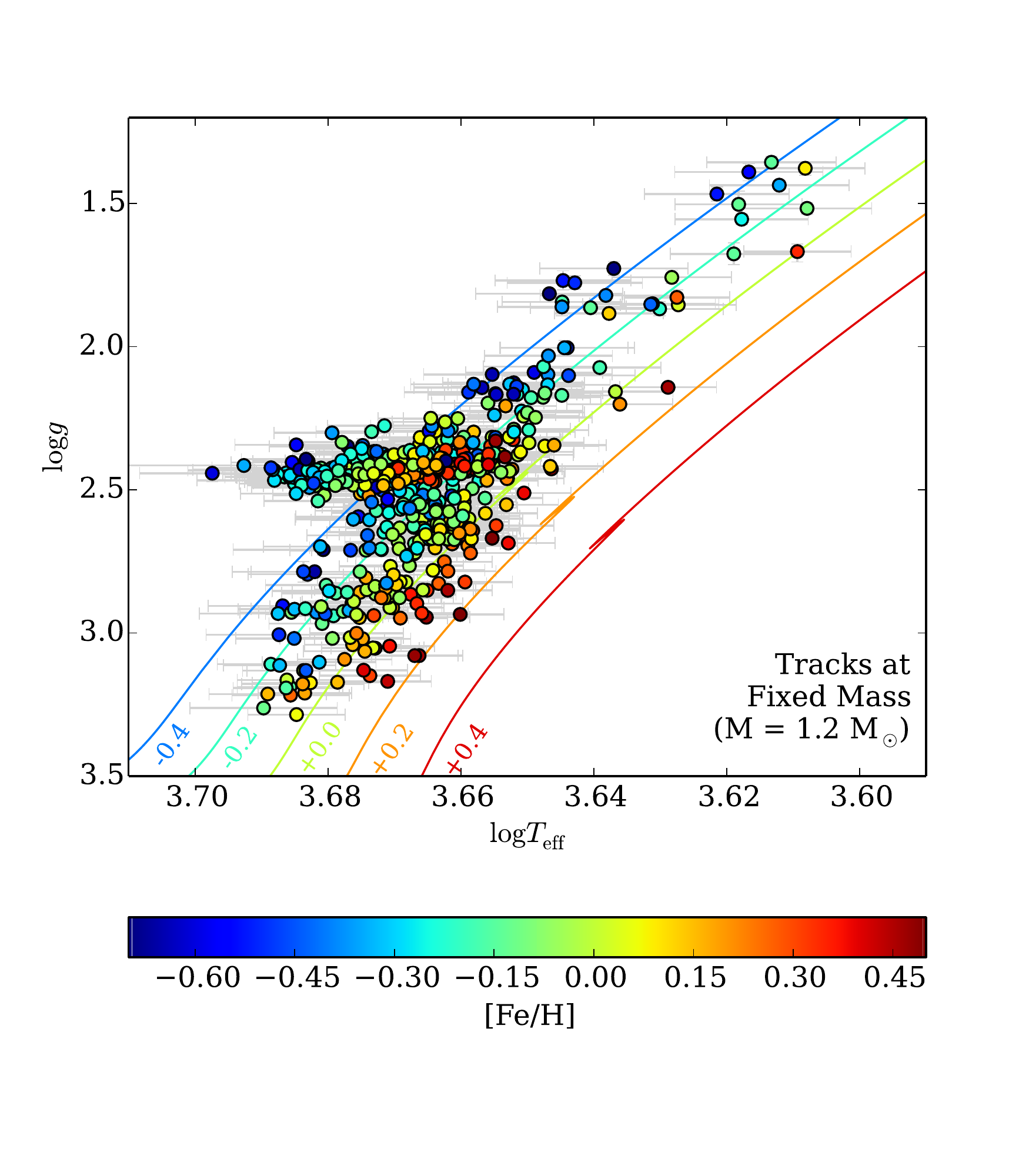}{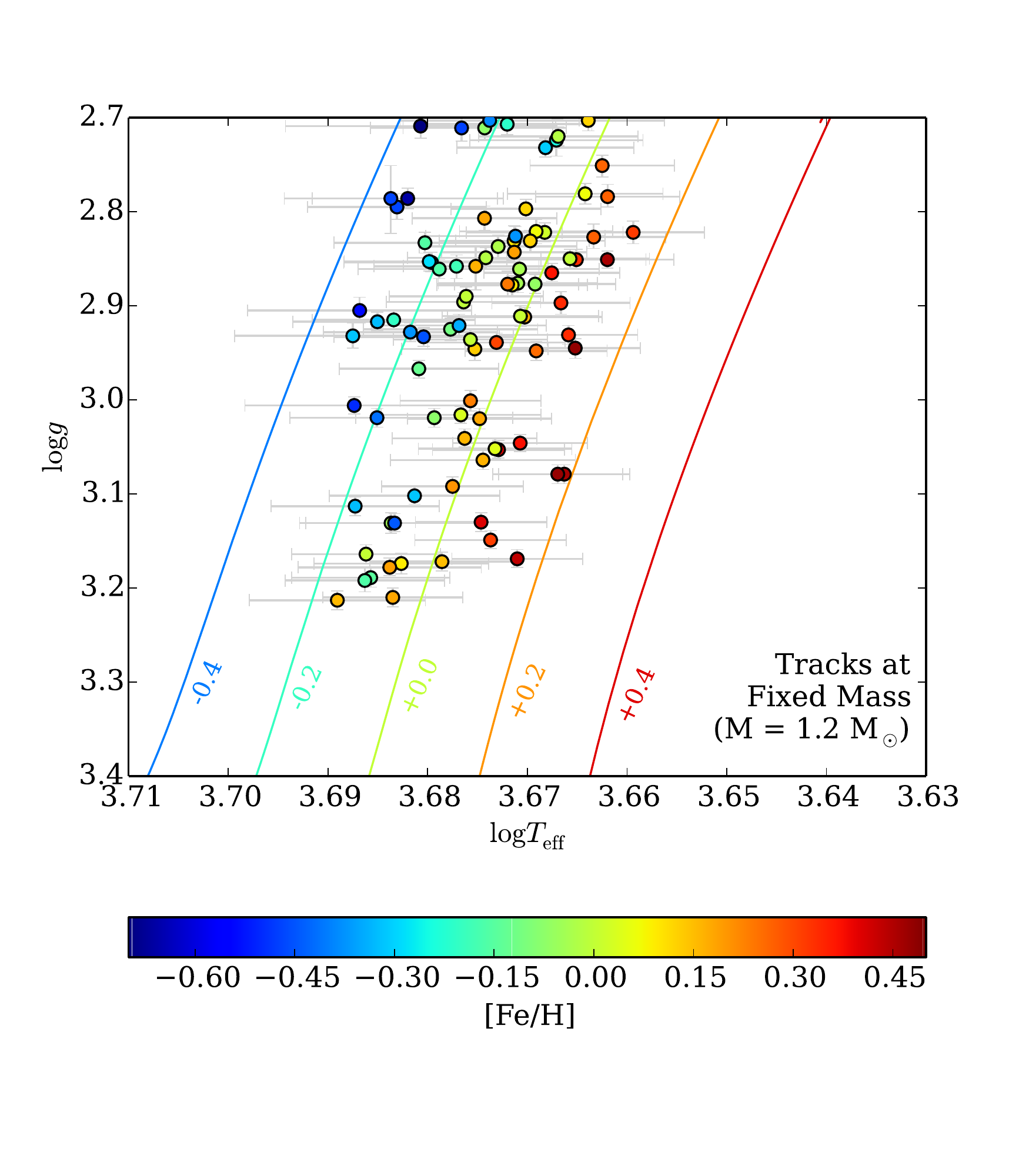}
  \caption{Left: HR diagram for all APOKASC stars in the mass bin $1.1<$ M $ < 1.3$ M$_\odot$, color coded 
by corrected ASPCAP metallicity such that lighter colors correspond to more metal-poor stars. For comparison, 
YREC tracks with 1.2 M$_\odot$ and [Fe/H]=-0.4, 0.0, and +0.4 dex are shown. Right: Same as for the left panel, 
except that the sample is restricted to seismically classified red giant branch stars on the lower red giant branch.
  There  is some evidence for a systematic offset in the magnitude of the 
metallicity trend for the highest metallicity targets.}
 \label{fig:MassTrends}
 \end{figure*}

We can also examine trends in mass at fixed metallicity (illustrated in Figure~\ref{fig:FeHTrends})
and in metallicity at fixed mass (illustrated in Figure ~\ref{fig:MassTrends}).  The overall trends predicted by
stellar models are clearly seen, with metal-poor stars being systematically hotter at 
fixed $\log g$ and mass than metal-rich ones.  Similarly, higher mass stars, as expected, are systematically
hotter at fixed metallicity than lower mass stars.  The agreement between theory and data degrades
for the lowest mass stars, which could imply a mass-dependent shift in the locus of the giant branch.
However, this may simply be a stellar population effect: the limited age of the disk
places a hard lower bound on the true mass of evolved red giants.  Stars with formal mass estimates at or
close to this value are likely to be higher mass stars scattered to a low apparent mass by errors in their
data, and they will thus appear offset relative to expectations.  For the YREC models there is good agreement
at low metallicity but there appears to be an offset in the temperature locus between the metal-rich track and
the metal-rich data.  This result is more robust than that in the mass plane and is tentative evidence 
for a metallicity-dependent offset between the expected and observed
HR diagram position at fixed mass.  \citet{Thygesen12} reported evidence for a metallicity (but not mass) dependent 
shift in the HR diagram position of 
stars with asteroseismic masses and gravities relative to theoretical expectations.  Our results are consistent with those that they derived from their sample.


\section{Conclusions, Cautions, and Future Prospects}

Asteroseismology and high-resolution spectroscopy are powerful and complementary astrophysical tools.  In this paper we present
the APOKASC catalog, which is the first large and homogenous catalog containing data from both high-resolution spectroscopy and asteroseismology.  
This effort provides new tools for studying stellar populations and testing the theories of both stellar atmospheres 
and stellar interiors. We now have a dataset measured in the natural co-ordinate for stellar evolution calculations, namely mass.  This permits direct tests
of stellar evolution tracks using isolated field stars, as opposed to relying on membership in well-studied binary systems or star clusters.  
Asteroseismic data is particularly precise and accurate for 
surface gravities, which is a stringent test of model
atmospheres theory.  Our combined data delineate stellar populations in the field with a precision usually associated with star clusters,
and asteroseismic evolutionary state diagnostics confirm the assignments (such as red clump, secondary red clump, or red giant branch) that would
have been expected from the HR diagram positions of the stars. Differential mass trends (at fixed metallicity) and metallicity trends (at fixed mass)
are in the expected sense, which is a powerful confirmation of expectations from stellar interiors theory.  At a deeper level, interesting trends 
in the differences between theory and data for both atmospheres and interiors models are clearly present and worthy of further exploration.

Typical uncertainties in our catalog are of order 80 K in $T_{\rm eff}$, 0.06 dex in [M/H], 0.014 dex in $\log g$, and $12 \%$
and $5 \%$  in mass and radius, respectively.  There are additional systematic error sources that could be important and which should be
explored.  For example, a zero-point shift in the $T_{\rm eff}$ scale is plausible given the differences between photometric and spectroscopic inferences
for our targets, which would induce correlated changes in the mass and radius values.  An improved extinction map for the \emph{Kepler} fields would permit
a more precisely anchoring of the absolute scale and constrain this effect.  Our metallicities show no evidence of systematic differences from targets measured with
optical spectroscopy.  We have derived mappings of KIC properties onto the system defined in this paper, and find a reasonable correspondence between the KIC and spectrscopic
metallicities for giants.  The accurate stellar properties and distances available in the APOKASC sample in the \emph{Kepler} field are also useful for calibrating stellar-population tracers, 
such that the power of the APOKASC sample can be leveraged to the large Galactic volume covered by the full APOGEE survey. As an example of this potential, \citet{Bovy14a} 
recently developed a new method for selecting RC stars from spectro-photometric data that is calibrated using the precise seismic $\log g$ and evolutionary-state 
classifications in the APOKASC catalog. The distances to stars in the resulting APOGEE-RC Catalog are accurate to $\sim\!5\,\%$, allowing for a detailed mapping of 
the structure of the Galactic disk. The seismic data are also useful to check for systematic errors in the RC distance scale, which is calibrated using \emph{Hipparcos} 
parallaxes: the direct seismic distances for 593 RC stars in common between the APOGEE-RC catalog and the Rodrigues et al. (2014) distance catalog agree to better than $1\,\%$.

At the same time, there are important factors that should be accounted for when interpreting and using the data.
There are significant sample selection effects, imposed by the target selection process for both \emph{Kepler} and APOGEE.  These selection 
effects are not simple ones (for example, magnitude or color cuts), and they reflect the fact that the asteroseismic data was obtained from a survey designed
for a very different purpose (detecting extrasolar planets.)  Direct stellar
populations inferences from our targets therefore require careful population modeling, which is outside the scope of our paper.

The availability of stellar masses is especially valuable for red giant stars, as the natural process of stellar evolution
channels stars of very different masses and ages into similar locations on the HR diagram.  The two major methods for inferring stellar masses also have significant limitations when applied 
to evolved stars, making an additional mass diagnostic even more important.  Direct mass measurements from binary stars are
uncommon for physically large red giant stars.  Indirect mass estimates for red giants in star clusters are subject to significant systematic uncertainties in cluster ages, distances, extinctions, 
and the mapping from turnoff properties to stellar mass.  By the same token, however, our asteroseismic masses have only a limited set of calibrators.  Our radii are based on the same scaling relations
used in \citet{SilvaAguirre12} and \citet{Huber12}, which reported good agreement
with independent tests using stars with measured parallaxes and interferometric angular size measurements, particularly at solar metallicity.  However, data at higher and lower metallicity are more
limited, and there could be metallicity-dependent offsets in the mass scale.  \citet{Epstein14}, for example, found that the masses derived from scaling relations for metal-poor stars were consistently
higher than those expected from other astrophysical constraints.  \citet{Miglio12} found evolutionary-state dependent differences between the radii inferred for red clump and red giant branch stars in
the metal-rich open cluster NGC 6791.  As a result, there could be systematic trends in our inferred masses and radii not captured in our analysis.  There are systematic differences between the
spectroscopic and asteroseismic surface gravities that appear to be a function of evolutionary state as well.  Calibration of the asteroseismic masses and radii, and understanding the origin of the trends that
have been identified, is ongoing.

Work is in progress on the next combined dataset, which will feature much larger numbers of targets and considerably broader phase space coverage in metallicity and
surface gravity.  Our asteroseismic uncertainties reflect a combination of systematic
and random error sources, and work is in progress to assess both separately.  We also adopted separate external calibrators for each of our spectroscopic parameters;
in the next version of the catalog we will explore the consequences of iterating between the asteroseismic and spectroscopic parameters and exploring the
impact of adopting an asteroseismic gravity prior for the spectroscopic solution.   In the current paper we employed a  
conservative concordance criterion
for our asteroseismic sample (see Section 4.1), reporting results only in cases where different analysis methods applied to the same dataset yielded similar asteroseismic parameters.
This approach had the net effect of removing most of the highest and lowest gravity targets from our sample, as their data is the most subject to systematic offsets.  
Including these in the next catalog, which will substantially increase our dynamic range in surface gravity, is another priority.  We also plan to take full
advantage of the improvements in the ASPCAP spectroscopic pipeline.

Funding for SDSS-III has been provided by the Alfred P. Sloan Foundation, the Participating Institutions, the National Science Foundation, and the U.S. 
Department of Energy Office of Science. The SDSS-III Web site is http://www.sdss3.org/. SDSS-III is managed by the Astrophysical Research Consortium 
for the Participating Institutions of the SDSS-III Collaboration including the University of Arizona, the Brazilian Participation Group, Brookhaven 
National Laboratory, Carnegie Mellon University, University of Florida, the French Participation Group, the German Participation Group, Harvard University, 
the Instituto de Astrofisica de Canarias, the Michigan State/Notre Dame/JINA Participation Group, Johns Hopkins University, Lawrence Berkeley National 
Laboratory, Max Planck Institute for Astrophysics, Max Planck Institute for Extraterrestrial Physics, New Mexico State University, New York University, 
Ohio State University, Pennsylvania State University, University of Portsmouth, Princeton University, the Spanish Participation Group, University of Tokyo, 
University of Utah, Vanderbilt University, University of Virginia, University of Washington, and Yale University.

CE, JJ, MP and JT would like to acknowledge support from NSF Grant AST-1211673.  DH acknowledges support by an appointment to the NASA Postdoctoral Program at
Ames Research Center administered by Oak Ridge Associated Universities, and NASA Grant NNX14AB92G issued through the Kepler Participating Scientist Program.
DS acknowledges support from the Australian Research Council. SB acknowldges partial support from NSF grant AST-1105930 and NASA grant NNX13AE70G.
Funding for the Stellar Astrophysics Center is provided by The Danish National Research Foundation (Grant agreement no.: DNRF106).  VSA was suppoeted by the European Research Council (Grant agreement no.: 267864).
SM acknowledges support from the NASA grant NNX12AE17G.  SH has received funding from the European Research Council under the European Community's Seventh Framework Programme (FP7/2007-2013) / ERC 
grant agreement no 338251 (StellarAges) and by Deutsche Forschungsgemeinschaft (DFG) under grant SFB 963/1 ``Astrophysical flow instabilities and turbulence''.
WJC, YE, and AM acknowledge support from the UK Science and Technology Facilities Council (STFC). D.A. acknowledges support provided by the National Research Foundation of Korea to the 
Center for Galaxy Evolution Research (No. 2010-0027910). AS is partially supported by the MICINN grant AYA2011-24704 and by the ESF EUROCORES Program EuroGENESIS (MICINN grant EUI2009-04170).
TCB acknowledges partial support for this work by grant PHY 08-22648: Physics Frontiers Center/Joint Institute for Nuclear Astrophysics (JINA), awarded by the U.S. National Science Foundation. We thank
an anonymous referee for comments that improved the paper.


\bibliographystyle{apj}

\end{document}